\documentclass[useAMS,usenatbib]{mn2e}
\pdfoutput=1
\usepackage[pdftex]{graphicx}
\usepackage{natbib}
\providecommand{\adsurl}[1]{{}}

\newcommand{\msun}{$M_{\odot}$}

\title[Moving Mesh Cosmology: Gas Accretion]{Moving mesh cosmology: tracing cosmological gas accretion}
\author[D. R. Nelson et al.]{Dylan Nelson$^{1}$\thanks{E-mail:
dnelson@cfa.harvard.edu}, Mark Vogelsberger$^{1}$, Shy Genel$^{1}$, Debora Sijacki$^{1}$\thanks{Hubble Fellow.}, Du{\v s}an Kere{\v s}$^{2}$\newauthor Volker Springel$^{3,4}$, Lars Hernquist$^{1}$\\\\
$^{1}$Harvard-Smithsonian Center for Astrophysics, 60 Garden Street, Cambridge, MA, 02138, USA\\
$^{2}$Department of Physics, University of California San Diego, 9500 Gilman Drive, La Jolla, CA 92093, USA\\
$^{3}$Heidelberg Institute for Theoretical Studies, Schloss-Wolfsbrunnenweg 35, 69118 Heidelberg, Germany\\
$^{4}$Zentrum f\"{u}r Astronomie der Universit\"{a}t Heidelberg, ARI, M\"{o}nchhofstr. 12-14, 69120 Heidelberg, Germany\\
}

\begin{document}

\maketitle

\begin{abstract}
We investigate the nature of gas accretion onto haloes and galaxies at $z=2$ using cosmological hydrodynamic simulations run with the moving mesh code {\small AREPO}. Implementing a Monte Carlo tracer particle scheme to determine the origin and thermodynamic history of accreting gas, we make quantitative comparisons to an otherwise identical simulation run with the smoothed particle hydrodynamics (SPH) code {\small GADGET-3}.
Contrasting these two numerical approaches, we find significant physical differences in the thermodynamic history of accreted gas in massive haloes above $\simeq 10^{10.5}$\msun. In agreement with previous work, {\small GADGET} simulations show a cold fraction near unity for galaxies forming in massive haloes, implying that only a small percentage of accreted gas heats to an appreciable fraction of the virial temperature during accretion. The same galaxies in {\small AREPO} show a much lower cold fraction, for instance $< 20\%$ in haloes with $M_{\rm halo} \simeq 10^{11}$\msun. This results from a hot gas accretion rate which, at this same halo mass, is an order of magnitude larger than with {\small GADGET}, together with a cold accretion rate which is lower by a factor of two. These discrepancies increase for more massive systems, and we explain both trends in terms of numerical inaccuracies with the standard formulation of SPH. We note, however, that changes in the treatment of ISM physics -- feedback, in particular -- could modify the observed differences between codes as well as the relative importance of different accretion modes.
We explore these differences by evaluating several ways of measuring a cold mode of accretion. As in previous work, the maximum past temperature of gas is compared to either a constant threshold value or some fraction of the virial temperature of each parent halo. We find that the relatively sharp transition from cold to hot mode dominated accretion at halo masses of $\simeq 10^{11}$\msun\, is a consequence of the constant temperature criterion, which can only separate virialised gas above some minimum halo mass. 
Examining the spatial distribution of accreting gas, we find that the filamentary geometry of accreting gas near the virial radius is a common feature in massive haloes above $\simeq 10^{11.5}$\msun. Gas filaments in {\small GADGET}, however, tend to remain collimated and flow coherently to small radii, or artificially fragment and form a large number of purely numerical ``blobs''. These same filamentary gas streams in {\small AREPO} show increased heating and disruption at 0.25-0.5 $r_{\rm vir}$ and contribute to the hot gas accretion rate in a manner distinct from classical cooling flows.
\end{abstract}

\begin{keywords}
cosmology: theory -- galaxies: evolution, formation, haloes -- methods: numerical
\end{keywords}


\section{Introduction}


The manner in which forming galaxies acquire their baryonic matter through cosmic time in a $\Lambda$CDM cosmology is not yet fully understood. In the classic picture of accretion from the intergalactic medium gas shock-heats to the virial temperature of the halo and forms a hot pressure supported atmosphere in approximate equilibrium \citep{rees77,silk77,wr78}. Gas can then cool onto a centrally forming galaxy, or not, based on the relevant cooling time-scale \citep{wf91}. In this regime the accretion rate onto the galaxy depends on the cooling rate and not on the cosmological accretion rate onto the halo itself. However, this holds true only for sufficiently massive haloes. Below some threshold mass the dark matter halo cannot support a stable hot, gaseous atmosphere \citep{bd03} and accreting gas does not necessarily shock heat to the virial temperature. This conclusion clearly motivates the differentiation of a ``cold accretion mode'' comprised of gas which does not shock heat to the virial temperature of its parent halo during accretion. Numerical simulations over the past decade have shown evidence for a cold accretion mode which dominates both at high redshift and for low mass objects at late times \citep{katz03,keres05,keres09,ocvirk08}. Haloes have been found to transition from cold to hot mode dominated at a mass in rough agreement with analytical estimates for the stable virial shock cutoff \citep[e.g.][]{bd03}.


Cold accretion in low mass objects below this threshold is precisely the regime where the virial shock is inactive. More massive haloes supporting a virial shock are also claimed to experience continuous, active cold mode accretion at high redshift ($z \geq 2$). This cold gas accretes along strongly filamentary geometries, termed ``cold flows,'' which penetrate the virial radius and deliver gas deep within the halo \citep{keres05,dekel09,db06}. In the context of analytical predictions, the non-spherical accretion of cold gas via highly collimated geometries was an unexpected result of these numerical simulations. Although 
smoothed particle hydrodynamics (SPH) simulations were the first to address the question of gas accretion modes in a cosmological context \citep{abadi03,keres05} they were soon followed by grid-based adaptive mesh refinement (AMR) efforts \citep[][and others]{ocvirk08}.


Simulations of cosmological volumes using the SPH technique typically make a cold mode gas selection based on the past temperature history of each gas element, invoking the pseudo-Lagrangian nature of the scheme. The resulting maximum past temperature $T_{\rm max}$ value is compared to some temperature threshold \citep{keres05,keres09,vdv11a,fgk11}. This threshold was originally motivated by the location of a minimum in the strongly bimodal global distribution of $T_{\rm max}$ found by \cite{keres05},
using the {\small TreeSPH} \citep{hk89} code. That work compared $T_{\rm max}$ to the virial temperature of haloes as well as to a constant temperature threshold, and chose $T_c \simeq 2.5 \times 10^5$\,K as sufficient to identify virial heating in haloes, due to the high halo masses considered. Essentially all SPH simulations to date have used this approach, although with less care as to its applicability, as we discuss later. One notable exception is \cite{brooks09} which used the SPH code {\small GASOLINE} \citep{wadsley04} and identified shocked gas particles based on an entropy jump criterion. This is a novel approach which should be explored in a full cosmological volume. The authors concluded that their method led to a selection of shocked/unshocked gas which agreed well with the selection of hot/cold gas based on the use of a constant temperature threshold.

More recently, finite volume techniques including cosmological re-simulations using adaptive mesh refinement (AMR) have been used to study the same problem. Grid-based simulations offer a better treatment of shocks, fluid instabilities, and phase boundaries. However for the analysis of cosmological gas accretion they are limited by their Eulerian nature, as the trajectory of any gas parcel cannot be simply 
identified due to mass exchange between cells. This difficulty can be overcome using tracer particles, and we argue these are a necessary feature of modern cosmological simulations. One AMR work to date \citep{dubois12a} used tracer particles in order to study the angular momentum of accreting gas. A common alternate approach is to use the instantaneous gas properties such as mass flux through a surface \citep[e.g.][]{ocvirk08, dekel09} to measure rates, or the low metallicity of gas as an indicator of cosmological origin \citep{agertz09}. Unfortunately these techniques cannot constrain the past thermal history of any gas element, nor identify for example material tidally stripped from a galaxy or subhalo during a previous strong interaction. Given these caveats, however, AMR simulations of single haloes have also found strong support for the importance of a filamentary, cold accretion mode in massive systems at high redshift.


The contribution of ``clumpy'' accretion due to infalling satellites and other bound substructures is also difficult to distinguish in Eulerian simulations. Some SPH studies use a stringent definition based on lack of membership to any identified bound substructure at any time prior to accretion \citep[e.g.][]{brooks09}. Others only require non-membership at some particular previous time \citep{keres05,vdv11a}. In either case, this gas 
should not be considered as contributing to cold accretion from the intergalactic medium (IGM). This is appropriate since we wish to isolate cosmological gas accretion from the IGM and not hierarchical buildup via mergers. The technical distinction is not particularly important as smooth accretion (defined as all gas not identified as clumpy) is generally found to be the dominant contribution to the total accretion rate \citep{vdv11a}. However, \cite{shen12} for example includes a substantial $\simeq$35$\%$ contribution of infalling dwarf satellites to the cold accretion mass budget. This effectively combines gas from multiple sources under a single label and ignores the diverse physical processes driving distinct accretion mechanisms. This is especially important in the results we report in what follows using {\small AREPO}, for which the accretion rate of cold gas directly from the IGM is small.


Only recently have simulations begun to explore the effects of feedback and galactic scale outflows on the character of gas inflow \citep{opp10,fgk11,vdv11a}. Differences in morphology between different modes of accretion in massive haloes and the added complexity from the physics of star formation implies that accretion rates may be quite sensitive to feedback processes. Simulations using SPH have concluded that a filamentary cold mode is largely unaffected by the presence of strong outflows arising from blastwave supernova feedback  \citep{shen12}. \cite{vdv11b} studied the impact of AGN feedback on inflow and concluded that it preferentially prevented hot mode gas (under the standard definition) from cooling from the halo onto the galaxy, while \cite{dubois12b} found that AGN feedback morphologically disturbed cold filaments. Current treatments of feedback in cosmological simulations, however, are largely phenomenological and are not self-consistent. Furthermore, the same numerical issues explored in \cite{vog12}, \cite{sijacki12}, \cite{keres12} and discussed herein that compromise the accuracy of SPH studies of gas inflow will also affect the interaction of outflowing ejecta and wind material with both halo and filamentary gas.  Moreover, the accretion rates onto galaxies are significantly higher in
simulations with {\small AREPO} than with {\small GADGET}, implying that feedback effects need to be even more efficient than in previous simulations with SPH.

Strong feedback may also indirectly affect the characterisation of inflows. In particular, galactic fountain material and other recycled gas may be a significant additional mode of accretion onto galaxies. \cite{opp10} included a phenomenological galactic wind model in cosmological SPH simulations and concluded that recycled gas accretion is a prominent ``third'' mode which in fact is the dominant accretion mechanism at $z \leq 1$, with minimal effect on this high redshift accretion with choice of outflow velocity. The ejection of hot gas from galaxies might also physically be expected to add mass and thermal support to a halo atmosphere. The result would be increased heating of cosmologically inflowing gas, an underestimated effect in all simulations lacking galactic scale outflows, including those presented here.


This paper presents results from a comparison of cosmological simulations run with the SPH code {\small GADGET} \citep{spr05b} and the moving mesh code {\small AREPO} \citep{spr10}. We compare individual objects as well as statistics derived from cosmological volumes and show that the importance and character of cold gas accretion depends sensitively on both analysis methodology and numerical technique. In Section \ref{sMethods} we describe the numerical details of our approach, two complementary tracer particle schemes, and our analysis methodology. Section \ref{sResults} compares the thermodynamics of gas accretion between the two codes,  discusses the various definitions of the cold mode and the sensitivity of our conclusions to those definitions, and investigates the geometry and filamentary nature of inflowing gas. Finally, Section \ref{sDiscussion} wraps up our discussion and summarizes the main conclusions.


\section{Methods} \label{sMethods}

\begin{table*}
  \caption{Details on the three different simulation resolutions and relevant numerical parameters.}
  \label{tbl_sims}
  \begin{center}
    \begin{tabular}{ccccccc}
     \hline 
     Gas Elements & DM Particles & Vel Tracers & MC Tracers & $m_{\rm target/SPH}$ [$h^{-1}$\msun] & $m_{\rm DM}$ [$h^{-1}$\msun] & $\epsilon$ [$h^{-1}$ kpc] \\ \hline\hline
     $128^3$ & $128^3$ & 1 x $128^3$ & 10 x $128^3$ & 4.8 x 10$^7$  & 2.4 x 10$^8$  & 4.0 \rule{0pt}{2.0ex} \\
     $256^3$ & $256^3$ & 1 x $256^3$ & 10 x $256^3$ & 6.0 x 10$^6$  & 3.0 x 10$^7$  & 2.0 \rule{0pt}{2.0ex} \\
     $512^3$ & $512^3$ & 1 x $512^3$ & 10 x $512^3$ & 7.4 x 10$^5$  & 3.7 x 10$^6$  & 1.0 \rule{0pt}{2.0ex} \\ \hline
    \end{tabular}
  \end{center}
\end{table*}

We use and extend the simulations and numerical methods presented extensively in \cite{vog12} and refer the reader to that paper for additional details. Brief descriptions of the key hydrodynamic differences between the two codes are presented here.

{\small GADGET} \citep[last described in][]{spr05b} is a smoothed particle hydrodynamics (SPH) implementation where the Euler equations are solved to model fluid flow by discretising mass using a set of particles. Continuous fluid quantities are defined by an interpolation kernel over a set of nearest neighbors. This method is pseudo-Lagrangian in nature (see discussion) and is particularly 
well-suited to handle the large dynamic range present in cosmological simulations where it is desirable to resolve both large scale structure and the internal dynamics of forming galaxies. We use a ``standard'' SPH formulation in {\small GADGET} to facilitate comparison to earlier work \citep{mon92} - the cubic spline kernel with $32$ neighbors and the density based formulation \citep{spr02} modified via an artificial viscosity $\alpha_{\rm SPH}=1.0$ with the Balsara switch \citep{bal95}.

{\small AREPO} \citep[described in][]{spr10} is a finite volume scheme where the control volumes are defined by a Voronoi tessellation of space. Euler's equations are solved using Godunov's method with the MUSCL-Hancock scheme to compute numerical fluxes and obtain second order accuracy. The tessellation is obtained by a set of mesh generating points which are allowed to move arbitrarily, though in our simulations follow the flow in a quasi-Lagrangian fashion. As a result {\small AREPO} retains the principal strengths of SPH including its adaptivity to a large dynamic range in spatial scales, Galilean invariance of the truncation error, and an accurate and efficient gravity solver \citep[e.g.][]{oshea05}. It also gains the strengths of finite volume codes, including the improved treatment of fluid instabilities, weak shocks (which can be missed in SPH; \citep[e.g.][]{keshet03}), 
phase interfaces, and shearing flows. We use the maximum cell face angle regularisation scheme and allow dynamic refinement and de-refinement to maintain approximately constant mass cells \citep{vog12}.

\subsection{Simulation Set} \label{ssSimSet}

All our simulations employ a WMAP-7 cosmology ($\Omega_{\Lambda,0}=0.73$, $\Omega_{m,0}=0.27$, $\Omega_{b,0}=0.045$, $\sigma_8=0.8$ and $h=0.7$) with a 20$h^{-1}$ Mpc sidelength volume at several different numerical resolutions, initially with equal numbers of dark matter particles and gas elements. The particle counts, initial gas element masses, dark matter particle masses, and Plummer equivalent gravitational softening lengths are given in Table (\ref{tbl_sims}). Unless otherwise noted, all results in this paper compare results between the $512^3$ resolution runs at a redshift of $z=2$.

{\small GADGET} and {\small AREPO} simulations include identical physics. We account for optically thin radiative cooling assuming a primordial H/He ratio which sets an effective temperature floor for gas in these simulations just below $\sim 10^4$ K \citep{katz96}. We do not include metal line cooling. A redshift-dependent ionizing UV background field \citep{fg09} is included as a spatially uniform heating source. Star formation and the associated ISM pressurisation from unresolved feedback events are included following \cite{spr03}. Gas elements are stochastically converted into star particles of a constant mass when the local gas density exceeds a threshold value of $n_{\rm H}=0.13$ cm$^{-3}$.

We do not include any explicit forms of strong stellar feedback that would drive galactic-scale winds. There is no treatment of black holes, radiative transfer and its effects, or magnetic fields. The key point is that we use a ``standard physics'' implementation which has been well studied and probes the interaction of gravity, hydrodynamics, and star formation in the cosmological context. The gravity implementation as well as the chosen subgrid physics are identical between our SPH and moving mesh simulations, allowing us to attribute differences in the solutions to differences in the method used to solve the equations of hydrodynamics. They are also similar to \cite{keres09}, in terms of the code and physics employed as well as the simulation resolution, enabling us to make a direct comparison to that work.

\subsection{Tracer Particles} \label{ssTracers}

In order to trace the evolution of gas properties over time in our moving mesh calculations we use a new ``Monte Carlo'' tracer particle technique (see Vogelsberger et al. and Genel et al., in prep). This probabilistic method associates tracers with parent gas cells and exchanges them based explicitly on the mass fluxes through each face. Specifically, all cells start with an equal number of tracers, after which each child tracer transfers from its original cell ($i$) to a neighboring cell ($j$) with a probability equal to $\Delta M_{ij} / M_i$, the ratio of the mass flux through face ($ij$) to the current mass of the originating cell. By construction, the tracer density is guaranteed to follow the underlying fluid density, at the cost of Poisson noise due to the probabilistic nature of the scheme. Furthermore, at each computational timestep each tracer records several fluid quantities of its parent gas cell - for this work we use only the maximum previous temperature and time of that $T_{\rm max}$ event. We initialize each gas cell with $10$ Monte Carlo tracers. Due to the target mass refinement scheme the number of child tracers per parent cell remains roughly constant throughout the simulation, with a Gaussian distribution extending a factor of two above and below the initial number. In particular, in galaxies at $z=2$ the mean number of child tracers is $\simeq$10 with a standard deviation of $\simeq$3.

We also evaluated a ``velocity field'' tracer particle scheme (see Vogelsberger et al. and Genel et al., in prep). In this more typical approach \citep[e.g.][]{vazza10,seit10,dubois12a}, tracers represent massless, passive particles which are purely advected by the local velocity field of the fluid. Our testing shows that this approach exhibits a systematic bias in its Lagrangian ability to follow mass fluxes. We suggest that the technique does not accurately recover the flow of mass in astrophysical situations involving convergent flows, including cosmological cases where the problem manifests as a ``pile-up'' of tracers in the centers of dark matter haloes (see Genel et al., in prep). The Monte Carlo approach is unaffected by this bias seen with the velocity field tracer particles. Furthermore, the flux-minimising moving mesh scheme allows us to use a reasonably small number of such tracers while achieving an acceptable level of noise (Genel et al., in prep). All results presented herein which involve tracking gas properties through time in the {\small AREPO} runs use our Monte Carlo tracer method. We however checked that our conclusions are qualitatively robust against choice of tracer scheme.

\subsection{Identifying Haloes} \label{ssIdentHaloes}

We identify dark matter haloes using the {\small SUBFIND} algorithm \citep{spr01} which begins with a friends-of-friends procedure (linking length $b=0.2$) applied to dark matter particles. Gas and stars are associated to their nearest dark matter particle. Tracers are naturally associated with their nearest (parent) gas cells. Next, an iterative unbinding procedure which accounts for thermal energy identifies substructures within each FOF group which are gravitationally bound groups, with a minimum of $20$ particles each. We refer to the largest such subgroup as the halo itself. Thus the halo selection specifically excludes gas in satellites and other orbiting substructures as well as unbound material, \textit{at the time of selection}. However, it is important to note that material associated but not bound to substructures, including tidal features such as tails, would not be explicitly excluded by this procedure. We discuss below the separate issue of classifying gas as either smooth or clumpy when following its thermal history.

The center of each halo is taken as the position of the particle with the minimum gravitational potential. We take as the virial radius $r_{\rm vir}$ the numerically computed $r_{\rm 200,crit}$\footnote{The radius within which the enclosed overdensity is 200 times the critical density $\rho_{\rm crit}$.}, and for the halo mass the mass of the largest bound subgroup. To determine accretion times, we need to estimate the time evolution of the virial radius of each halo as well as its position. However, {\small SUBFIND} is a purely 3D algorithm with no explicit time linking. We construct a parent tree (a simplified merger tree) by matching haloes between successive snapshots based on the largest fraction of common dark matter particles. We require $60\%$ particle agreement and reasonable mass and center offsets between snapshots. For consistency we restrict our analysis on a halo by halo basis to the time period for which the parent tree is well defined, which is true for $>$\,$90\%$ of haloes with $M_{\rm halo} \geq 10^{10}$\msun\, over the chosen time window, as discussed below.

\subsection{Maximum Past Temperature} \label{ssMaxPastTemp}

\begin{figure*}
\centerline{\includegraphics[angle=0,width=6.4in]{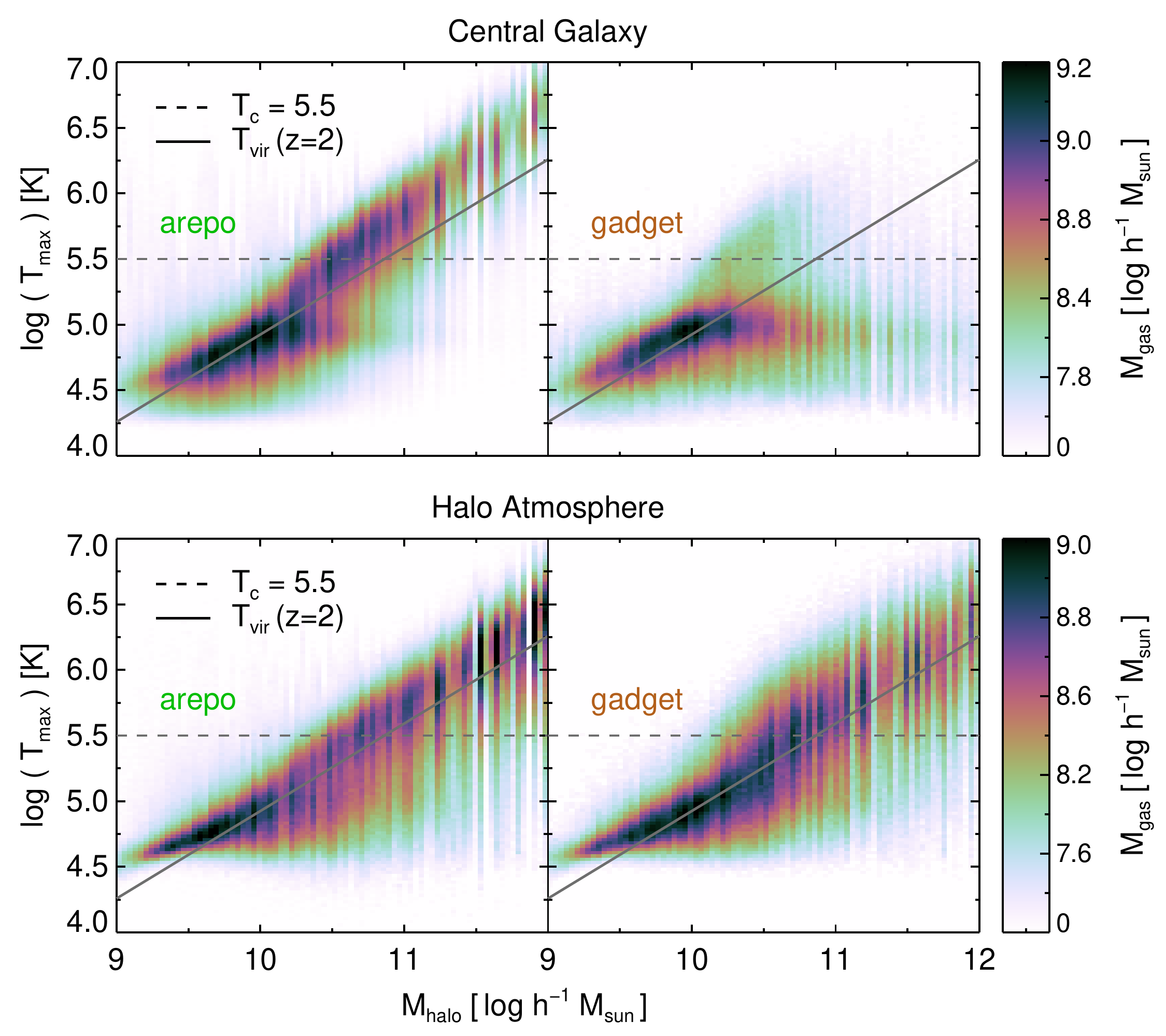}}
\caption{ The mass-weighted histogram of past maximum temperature $T_{\rm max}$ for gas accreted onto central galaxies (top) and halo atmospheres (bottom) by $z=2$, as a function of the parent halo mass at $z=2$. {\small AREPO} and {\small GADGET} are shown separately in the left and right panels, respectively. In this figure we do \textit{not} normalize these temperatures by any fraction of $T_{\rm vir,cur}$ or $T_{\rm vir,acc}$. A constant temperature of $T_c = 10^{5.5}$\,K is shown (dashed line) together with the virial temperature at redshift two (solid line). The two codes deviate strongly for galaxies in haloes with masses above $\simeq 10^{10.5}$\msun, where the strong contribution of hot gas in {\small AREPO}, which scales approximately with $T_{\rm vir}$, is mostly absent in {\small GADGET}. The cold $\sim 10^5$\,K gas which dominates the accretion in {\small GADGET} galaxies is largely absent in the {\small AREPO} simulation.
 \label{fig_tmax2d}} 
\end{figure*}

We broadly consider three different means of measuring cold mode accretion. Two depend on the properties of associated haloes. We introduce three important temperature values used throughout this work:

\begin{itemize}
\item $T_{\rm vir,cur}$ is the virial temperature for the parent halo of each tracer or gas particle at the ``current'' time (i.e. the time the gas selection is made, $z=2$).

\item $T_{\rm vir,acc}$ is the virial temperature evaluated at the ``accretion'' time on a particle by particle or tracer by tracer basis. We define the accretion time here as the most recent virial radius crossing time.

\item $T_c$ refers to a constant temperature threshold at some value, applied uniformly across all halo masses.
\end{itemize}

To estimate the virial temperature we use the definition of \cite{bl01} which at redshift two gives

\begin{equation}
T_{\rm vir} = \frac{\mu m_p V_c^2}{2 k_B} \simeq 4 \times 10^5 \,\rm{K} \left(\frac{\mathit{M}_{\rm halo}}{10^{11} \, \mathit{h}^{-1 } M_{\odot}}\right)^{2/3} \left(\frac{1+z}{3}\right)
\end{equation}

\noindent where $V_c$ is the circular velocity at the virial radius and $\mu \simeq 0.6$ for fully ionized, primordial gas.  We note that the redshift scaling as expressed here is only approximate, as it ignores small changes in cosmological parameters at the redshifts of interest.

We calculate a cold fraction for each halo or galaxy as the ratio of cold to hot accreted gas at $z=2$ using each of the three methods. To calculate the maximum past temperature of a gas selection in {\small GADGET} runs we examine the temperature of each SPH particle at each snapshot back to the start of the simulation. We use 314 snapshots logarithmically spaced in redshift from $z=30$ to $z=0$, corresponding to a mean spacing at $z=2$ of $\simeq 25$ Myr. To calculate the maximum past temperature of a gas selection using the Monte Carlo tracers in {\small AREPO} runs we first find all the tracer children of those gas cells at the target redshift. These tracers are then treated as a uniform population and the maximum past temperature of each, recorded for each active timestep of their parent gas cell, is obtained since the start of the simulation. That is, the mass weight of the cell is divided among its children tracers and each may potentially have a distinct thermodynamic history. In both cases we neglect temperature maxima while gas is on the effective equation of state (star forming).

\subsection{Measuring Accretion} \label{ssMeasAcc}

We separate gas in the ``central galaxy'' from the ``halo atmosphere'' by a cut in the density, temperature plane as

\begin{equation}
\log{( T_{\rm gas} [\rm{K}] )} - \frac{1}{4} \log{( \rho_{\rm gas} [10^{10} M_{\odot} \,h^2 \mathrm{kpc}^{-3}])} < 6.0
\end{equation}

\noindent following \cite{torrey12}. This selection differentiates between hot gas in the halo and the cold, dense gas which is rotationally supported at the center of the halo. We add a radial cut at $0.15 r_{\rm vir}$ to this separation criterion to prevent including any cold gas at large radii as part of the galaxy. Star particles are also included throughout our analysis and are considered as part of the galaxy if they satisfy the radial cut. Our results are quantitatively unaffected for reasonable choice of these three galaxy selection parameters.

In order to define accretion and measure accretion rates we introduce a ``time window'' over which accretion events are counted. A gas element is considered to have accreted onto a halo if it belonged to that halo at $z=2$ and crossed the virial radius during the chosen time window. A gas element is considered to have accreted onto a galaxy if it belonged to that galaxy at $z=2$ and either crossed the phase space cut in $(\rho,T)$ or the radial cut at $0.15 r_{\rm vir}$ during the chosen time window. The accretion time is taken to be the more recent (lower redshift) of these two events. The same definition is used for star particles in galaxies, where we search for the accretion event of the progenitor of each star particle only in the gas phase. That is, stars which remain stars for the duration of the chosen time window are not considered as accreted. Rates are calculated by normalising the total accreted mass (counting particles or tracers) by the time period.

All accreted material is separated into one of three disjoint ``modes'' of accretion based on the following criteria, evaluated at the virial radius crossing time for each gas element:

\begin{itemize}
\item Smooth - not a member of any halo or subhalo, other than the main progenitor branch of its parent halo, and likewise back to $z=6$.

\item Clumpy - gravitationally bound to a halo or subhalo, other than the main progenitor branch of its parent halo.

\item Stripped - not a member of any halo or subhalo (smooth), but gravitationally bound to some other halo or subhalo at any previous time back to $z=6$.
\end{itemize}

These definitions allow us to exclusively consider accretion from the IGM by restricting our analysis to the smooth mode only. This is required in order to remove the ``merger contribution'', and is similar in spirit to the approach taken by previous studies of cosmological gas accretion for which it is possible to make this distinction (SPH and not grid codes). However, we note that our definition of clumpy is somewhat more restrictive than in previous works \citep[e.g.][]{keres05} which would remove only the central galaxy component of each subhalo. By taking all gas bound to substructures as a merger contribution we impose a stronger restriction, and therefore remove a larger fraction of material in our analysis of smooth accretion. Finally, we measure the importance of tidally stripped material with the third mode, which is not direct accretion from the IGM but rather gas which has been previously gravitationally bound to some halo or galaxy. Unless otherwise noted, all results in this paper include only smoothly accreted material over an accretion time window of 1 Gyr. The only exceptions are Figure \ref{fig_totalmass} which considers the total accreted baryonic mass over an ``integrated'' time window extending back to $z=6$, and Figure \ref{fig_accRateMode} which compares the smooth, clumpy, and stripped accretion modes.


\section{Results} \label{sResults}

\begin{figure*}
\centerline{\includegraphics[angle=0,width=6.4in]{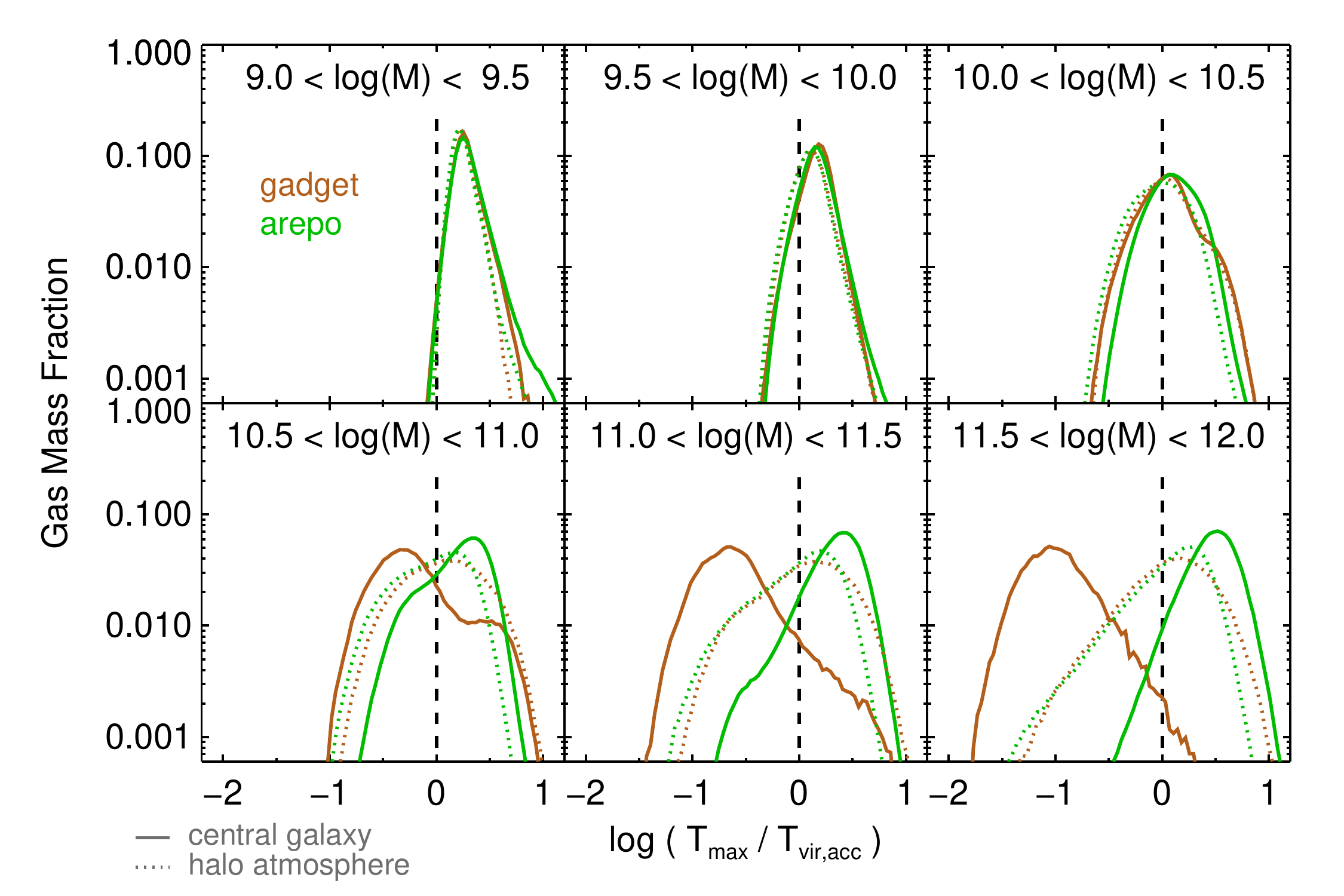}}
\caption{ Relative distributions of past maximum gas temperature normalized by the virial temperature of the parent halo at the time of accretion, without any cold or hot mode differentiation. Each of the four curves in each panel is separately normalized, so that only their positions and shapes, but not their amplitudes, should be compared. At low halo masses (top three panels) and for gas accreted onto haloes (dotted lines) the two codes agree well. For more massive haloes above $\simeq 10^{10.5}$\msun\, significant differences arise in the thermal history of gas accreted onto galaxies, where gas in {\small GADGET} has predominantly $T_{\rm max} < T_{\rm vir,acc}$ while gas in {\small AREPO} has predominantly $T_{\rm max} > T_{\rm vir,acc}$.
 \label{fig_tmax_tviracc}} 
\end{figure*}

In this section we investigate differences between {\small GADGET} and {\small AREPO} in the thermodynamic history of gas accreted onto haloes and the galaxies forming at their centers. Our primary proxy for this potentially complicated history is $T_{\rm max}$, the maximum past temperature of each gas element. In Figure \ref{fig_tmax2d}\ we show the mass-weighted histogram of $T_{\rm max}$ for smoothly accreted gas, as a function of parent halo mass. We also indicate a  nominal constant temperature threshold value along with the virial temperature at redshift two.

Firstly, we see that {\small GADGET} and {\small AREPO} agree well on the relative distribution of past maximum temperatures contributing to \textit{halo atmospheres} across the entire mass range (lower panels). For accretion onto galaxies (upper panels) at all halo masses the {\small AREPO} temperature distribution scales approximately with $T_{\rm vir}$, while in {\small GADGET} this behavior holds only for low mass haloes below $\simeq 10^{10}$\msun. In fact, we find that gas heating proceeds in relation to the virial temperature even for low mass systems.  At $\simeq 10^{9}$\msun\, the $T_{\rm max}$ distribution may be beginning to scale somewhat shallower than $T_{\rm vir}$, flattening towards a uniform, cold temperature set by the IGM of a few times $10^4$\,K. However, our simulation resolution is not sufficient at these low mass systems to convincingly demonstrate this. There is also an indication of a roughly constant temperature component at $T_{\rm max} \simeq 10^5$\,K in the {\small GADGET} simulation, which we discuss further below. The scaling with $T_{\rm vir}$, together with the lack of any distinguishable segregating feature about a constant temperature threshold on the order of a few times $10^5$\,K, motivates the comparison of $T_{\rm max}$ to $T_{\rm vir}$ on a halo by halo basis.

In Figure \ref{fig_tmax_tviracc}\ we show the distribution of $T_{\rm max}$ for smoothly accreted gas, onto both galaxies and haloes, divided into six bins of halo mass. In each case we now normalize temperatures by the virial temperature of the dark matter halo of each gas element, at the time of accretion. For accretion onto central galaxies for haloes with $M \leq 10^{10.5}$\msun\, the two codes show excellent agreement. For such low mass systems, however, this agreement does not necessarily indicate a similar level of shock heating. The virial temperatures of these haloes are sufficiently low that IGM pre-heating begins to set an effective temperature floor for the least massive systems. This heating outside the halo occurs through a combination of radiative energy input from the UV background, as well as shocks during large scale collapse, both of which we discuss further in the context of cold fractions. It is only for $M_{\rm halo} \geq 10^{10.5}$\msun\, where gas accreted onto centrally forming galaxies begins to show different thermal histories between {\small GADGET} and {\small AREPO}. At these masses, the majority of gas in the SPH calculation is never heated to temperatures comparable to the halo virial temperature at accretion. In contrast, a dominant fraction of the same accreted gas in {\small AREPO} reaches approximately the virial temperature, or above. This striking difference in the temperature history of gas accreted onto galaxies in haloes with $M_{\rm halo} \geq 10^{10.5}$\msun\, is the first main result of this paper.

In this same mass regime both codes show evidence for potentially bimodal temperature distributions. This is most evident for {\small GADGET} galaxies between $10.5 \le$ log(M) $\le 11.0$ where $T_{\rm max} / T_{\rm vir,acc}$ could be well described by the sum of two symmetric distributions with different peak temperatures. As normalized, the colder peak moves to lower temperatures with increasing halo masses approximately as $T \propto M^{2/3}$ and is consistent with a constant physical temperature of $T \simeq 10^{5}$\,K, as seen in Figure \ref{fig_tmax2d}. We note that common choices for a constant temperature threshold adopted by previous studies are above this value. The warmer peak shows no scaling with halo mass and is consistent with a heating process scaling with virial temperature. These two contributions represent, respectively, accretion of roughly constant temperature material from the IGM which has minimal interaction with the halo potential, and material which virialises to a mass dependent temperature. Results from both {\small GADGET} and {\small AREPO} are consistent with this picture, though they strongly disagree on the relative contributions of these two accretion channels.

\subsection{Accretion Rates}

\begin{figure*}
\centerline{\includegraphics[angle=0,width=5.8in]{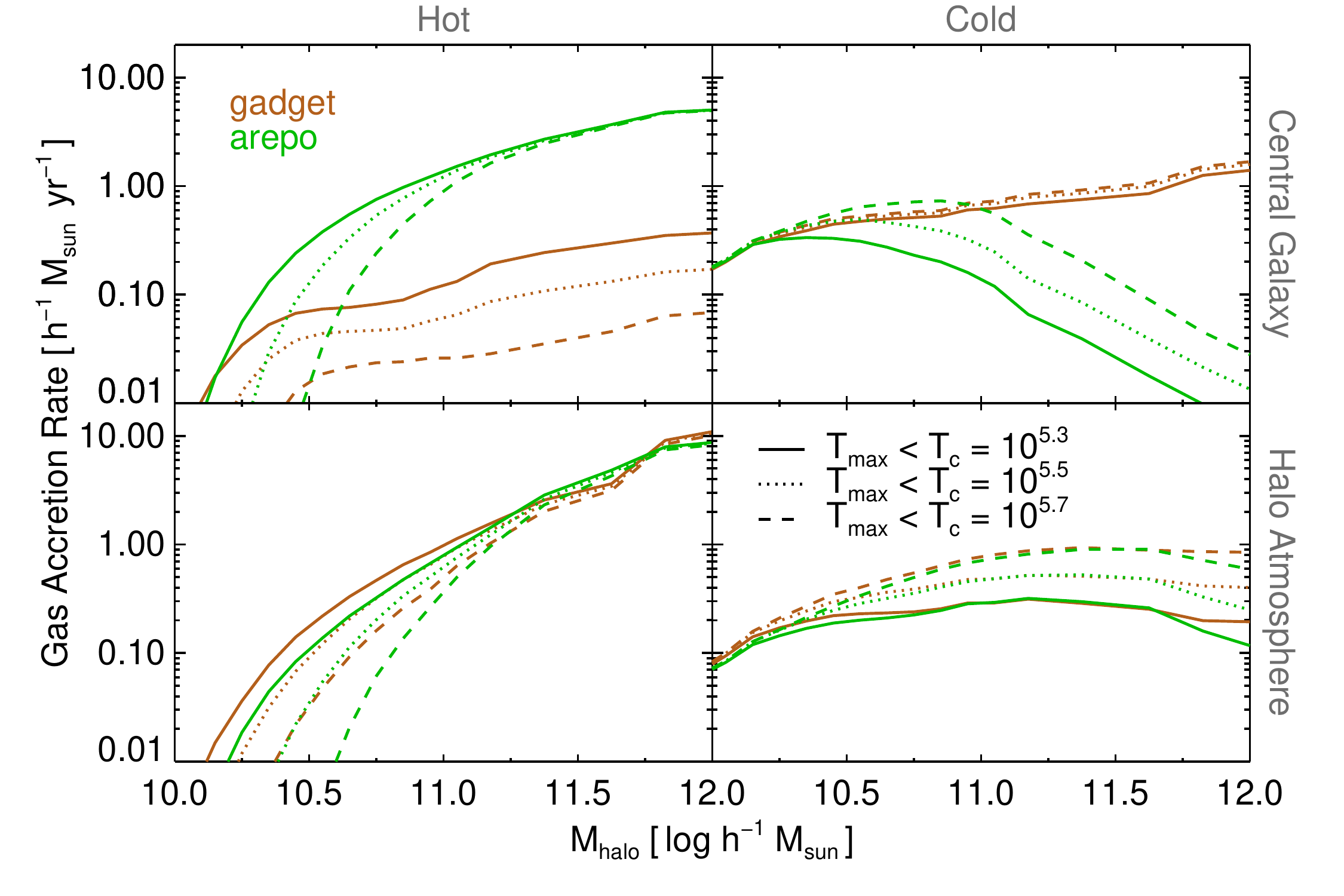}}
\caption{ Gas accretion rate onto galaxies (top) and haloes (bottom). In all four panels we define the cold mode as requiring $T_{\rm max}$ be less than a constant value $T_c$, as listed, demonstrating the sensitivity of the derived rates to the method of measuring the cold versus hot mode. 
 \label{fig_accrate1}} 
\end{figure*}

\begin{figure*}
\centerline{\includegraphics[angle=0,width=5.8in]{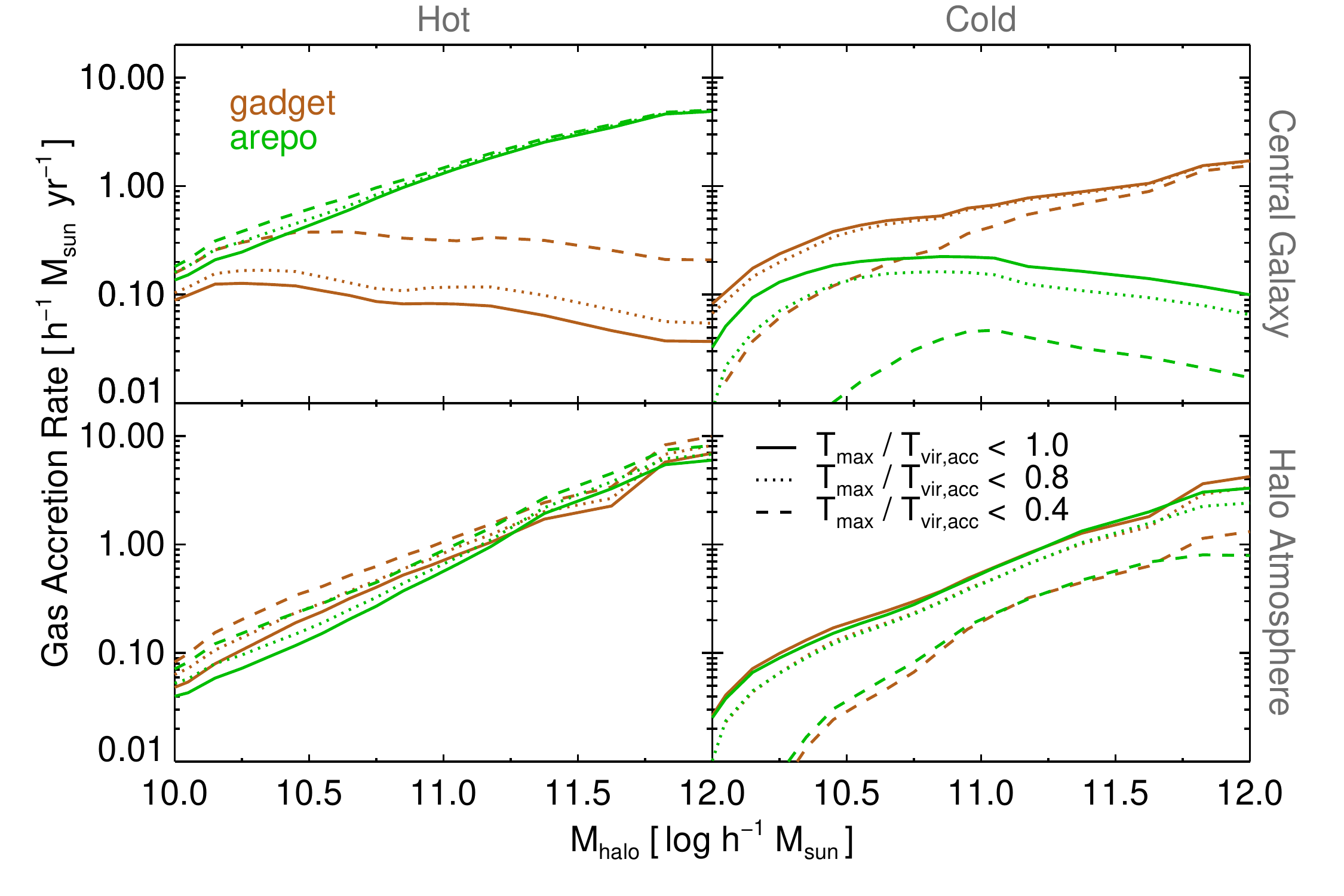}}
\caption{ As in the previous figure, exploring the sensitivity of the gas accretion rate onto galaxies (top) and haloes (bottom) to definition of the cold mode. Here we take cold mode gas as having $T_{\rm max} / T_{\rm vir,acc} < {1.0,0.8,0.4}$, as listed. 
 \label{fig_accrate2}} 
\end{figure*}

\begin{figure*}
\centerline{\includegraphics[angle=0,width=5.8in]{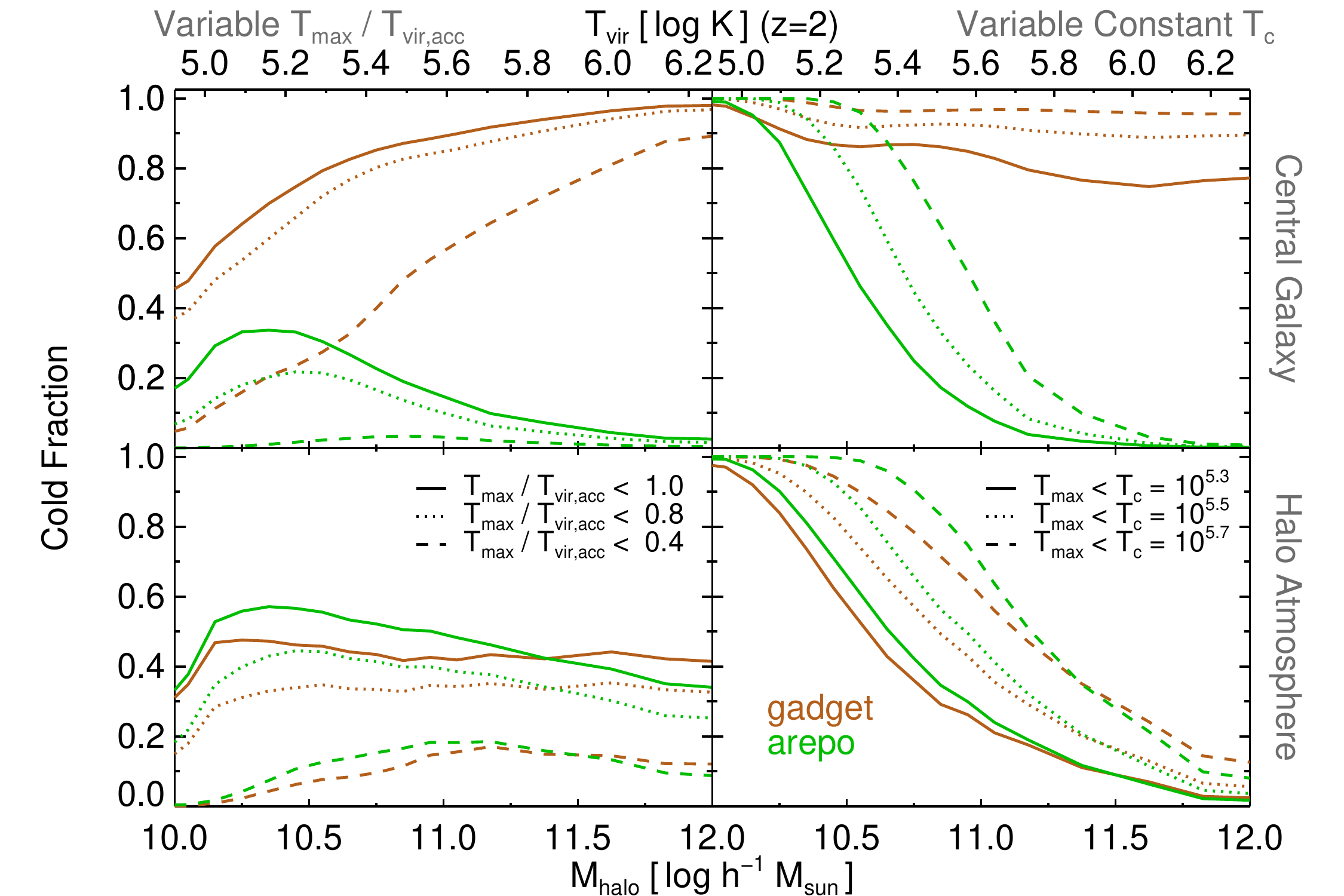}}
\caption{ Sensitivity of the derived cold fraction to the manner in which it is measured. The top two panels consider gas accreted onto central galaxies whereas the bottom two panels consider gas accreted onto halo atmospheres. The left two panels consider a gas element to have been accreted cold if the fraction $T_{\rm max} / T_{\rm vir,acc}$ is less than $\{1.0,0.8,0.4\}$. The right two panels define cold mode gas as having $\log {( T_{\rm max} [\rm{K}] )}$ less than a constant value of $\{5.7,5.5,5.3\}$. We find that the most massive systems which have $f_{\rm cold} \simeq 1$ in {\small GADGET} have a cold fraction approaching zero in {\small AREPO}, while the relatively sharp transition from cold to hot mode dominated, seen clearly in the constant temperature moving mesh case, is not evident when using the virial temperature comparison.
 \label{fig_coldfrac_var}} 
\end{figure*}

To understand the nature of the differences, we measure the accretion rate onto both central galaxies and haloes themselves as a function of halo mass. Figures \ref{fig_accrate1}\ and \ref{fig_accrate2}\ separate cold from hot mode accretion based on comparison of $T_{\rm max}$ to either a constant temperature threshold $T_c$ or the halo virial temperature $T_{\rm vir}$, respectively. The lower panels of both figures again indicate that the total accretion rates and thermal history of gas which accretes onto the halo itself is similar between the two codes. Our accretion rates for the low mass end are somewhat lower than in \cite{keres09} for instance, which may be due to our more stringent definition of smooth accretion as well as our better resolution of the hierarchical nature of structure growth. When using a $T_{\rm max} / T_{\rm vir,acc}$ comparison this scaling is well described by a power-law in the mass range where systems are both well resolved and sufficiently numerous. We now discuss how the accretion rates compare between the two codes based on Figure \ref{fig_accrate2}\ and $T_{\rm vir,acc}$, returning later to this particular choice.

The accretion rate onto galaxies in {\small GADGET} is dominated by gas which has always been relatively cold, across the entire mass range of $10^{10} \leq M_{\rm halo} \leq 10^{12}$. In {\small AREPO} the same galaxies are predominantly fed by gas which has experienced significant heating. This contrast is robust under any reasonable comparison of maximum past temperature to either $T_c$ or $T_{\rm vir}$. The accretion rate in the hot mode is different by roughly an order of magnitude at $M_{\rm halo} \simeq 10^{11}$\msun\, between the two codes, having essentially zero contribution to haloes above this mass in the SPH simulation. The primary reason for this difference is the efficient cooling of hot halo gas in {\small AREPO} which is artificially suppressed in {\small GADGET} by spurious viscous heating. \cite{bauer12} studied the dissipation of energy from subsonic turbulence in SPH and found that spurious heating arises both from the premature dissipation of turbulent energy on large spatial scales and the viscous damping of SPH-noise on small scales. In the context of our current cosmological simulations, this prevents gas at the cooling radius from forming a proper Kolmogorov-like spectrum. The viscous heating offsets part of the gas cooling, keeping the gas locked in the diffuse, hot phase. In {\small AREPO} this same turbulent energy is more realistically dissipated by cascading to smaller spatial scales and higher densities, allowing significantly more efficient cooling of hot gas in the atmospheres of massive haloes \citep{keres12}. This cooling channel is a dominant accretion mode of massive haloes in the moving mesh calculation.

The accretion rate of cold gas onto galaxies also differs between the two calculations. At $M_{\rm halo} \simeq 10^{11}$\msun\, {\small GADGET} galaxies have a factor of two larger cold accretion rates as corresponding {\small AREPO} galaxies, and this discrepancy grows with increasing halo mass to an order of magnitude effect for $M_{\rm halo} \simeq 10^{12}$. We identify two primary numerical issues that cause the observed artificial increase of the cold gas accretion rate in the SPH calculation. Both operate by increasing the efficiency with which gas is heated in various regimes, resulting in a redistribution of gas to higher temperatures in the {\small AREPO} haloes.

First, \cite{creasey11} demonstrate the issue of numerical overcooling due to shock broadening which arises due to the artificial viscosity treatment in SPH. At comparable resolution, the shock-capturing Godunov scheme in {\small AREPO} better resolves the resulting post-shock gas properties (including temperature), critical since smoothly accreting gas undergoes multiple small shocks as it virialises with hot halo gas. Secondly, spurious pressure forces in regions of steep density gradients \citep{agertz07}, particularly contact discontinuities, lead to the numerical fragmentation of gas structures into artificial ``blobs'' due to surface tension \citep[see][]{torrey12}. A discussion of the formation mechanism of these blobs and their prevalence in different formulations of SPH can be found in \cite{hobbs12}. Their presence is a significant contribution to the accretion rate of cold gas in {\small GADGET}. We have checked explicitly that gas in these blobs have sufficiently low $T_{\rm max}$ to be included in the cold accretion rate regardless of definition. For $M_{\rm halo} \geq 10^{11}$\msun\, these radially penetrating blobs contribute to the cold, smooth accretion rate in {\small GADGET}, directly transporting low temperature, low angular momentum material into the centers of haloes \citep[e.g.][]{kh09}. No comparable gas features are present in the moving mesh calculation, and this cold gas contribution to galactic accretion is entirely absent.

\subsection{Cold Fractions and the Transition Mass}

The ratio of total accreted mass with $T_{\rm max}$ below some temperature threshold to the total mass above -- the cold fraction, as a function of halo mass -- is a widely reported quantity in studies of cosmological gas accretion, dating back to \cite{keres05}. Figure \ref{fig_coldfrac_var}\ explores how robust the derived cold fractions are to different types of parameter choices in the measurement of a cold mode. Using a constant temperature comparison (upper right panel), we note that the cold fraction for galaxies derived from our {\small GADGET} simulations are in agreement with \cite{keres09}. As expected from our previous discussion of accretion rates, this cold fraction is near unity for all halo masses, while in {\small AREPO} the cold fraction drops to zero for galaxies hosted in sufficiently massive haloes above $\simeq 10^{11}$\msun. We have seen already that lower cold fractions in {\small AREPO} arise from both a lower total gas mass accreted through a cold channel, as well as a larger total gas mass accreted through a hot channel. 

We define the ``transition mass'' as the halo mass at which the cold fraction equals $1/2$. Using the constant temperature cut $T_c$ $\simeq$ 5.5 \citep{keres05} leads naturally to a transition from $f_{\rm cold}=1$ to $f_{\rm cold}=0$ at a halo mass of $\simeq 10^{11.0}$\,\msun in our highest resolution simulation. It is tempting to compare this to model predictions for the critical halo supporting a virial shock. \cite{bd03} showed that this transition mass, defined as the point where the cold and hot modes have equal contribution, scales due to the contribution of metals to the cooling curve from $\simeq 10^{10.9}$\msun\, at $Z=0$ to $\simeq 10^{11.6}$\msun\, at $Z=0.3$ (redshift two). However, we observe that the appearance and location of this transition is by construction due to the choice of $T_c$. Given the scenario where all haloes heat some large fraction of infalling gas to their virial temperature, this use of a constant temperature threshold would also reproduce a transition in this mass range due solely to the changing virial temperatures. Consider the case of a halo with a virial temperature just below the chosen constant temperature threshold, corresponding to a mass of $\simeq$10$^{11}$\msun\, taking the usual choice for $T_c$. Even if this halo shock heats all accreting gas to its virial temperature it would nevertheless be attributed a cold fraction of $100$\% under this method of measuring the cold accretion, which is clearly not a physically meaningful statement. We find that the subsequently derived transition mass increases monotonically with the chosen $T_{c}$ if a constant temperature criterion is used to separate the two modes. From $T_c=5.3$ to $T_c=5.7$ (a factor of $\simeq 2.5$) the transition mass increases by roughly half a decade, which is consistent with the $T^{3/2}$ scaling expected solely due to the increasing virial temperature.

If we instead use a virial temperature comparison (upper left panel), the derived cold fractions show markedly different behavior. In this case {\small GADGET} galaxies exhibit a monotonic increase in their cold fraction from $\simeq 0.4$ to $1.0$ from $M_{\rm halo} = 10^{10}$ to $10^{12}$\msun. Galaxies in low mass haloes below $M_{\rm halo} \simeq 10^{10}$\msun\, are no longer cold mode dominated, while those in high mass systems are. In contrast, {\small AREPO} galaxies still show the broad trend of decreasing cold fraction with increasing halo mass, though it tends towards continual evolution as opposed to sharp transition. Using the virial temperature as the dividing criterion on a halo by halo basis makes any  transition significantly broader in halo mass. The second key result of this paper is then, given this more physically motivated means of measuring cold mode accretion in lower mass haloes, there is no longer a clear feature in the cold fraction as a function of halo mass.

Although we argue for a comparison against $T_{\rm vir}$ and not $T_c$ in this work, it is important to note that the two methods work equally well within a specific halo mass range. In particular, for haloes with $T_{\rm vir} \gg T_c$ such that virialisation related heating is well separated by this cut. At $z=2$ this must require as a minimum $M_{\rm halo} \geq 10^{11}$\msun\, which is indeed the mass range considered in early studies \citep[e.g.][]{keres05}. Unfortunately this minimum halo mass falls well within the cold-hot transition range as reported at this redshift as well as at redshift zero, where the restriction becomes $M_{\rm halo} \geq 10^{12}$\msun\, -- the midpoint or even tail of the transition \citep{vdv11a,fgk11}. Caution must also be taken using any $T_{\rm vir}$ comparison as the more restrictive selections of the cold gas selection ($0.4,0.8$) for low mass haloes fall below the effective temperature floor of our simulations. We run a separate experiment at the $256^3$ resolution which excludes the heating contribution from the UV background. The resulting temperature distribution of IGM gas extends to significantly lower temperatures, and the turnover in cold fraction around halo masses of $\simeq$10$^{10.5}$\msun\, essentially disappears. We then attribute this feature in the cold fraction to the high IGM gas temperature prior to accretion. This also indicates the mass scale below which a temperature criterion of any kind will be unable to separate physically distinct accretion modes.

Finally, we note that switching from the ``current'' (not shown) to the ``accretion'' virial temperature criterion leads to a  $\sim 10\%$ change, lowering the cold fraction across all halo masses. This change is unlikely to have great impact on the general interpretation of the cold mode importance. However, the virial temperature of a particular halo can change significantly between the time material is accreted and $z=2$. To physically motivate our analysis requires that we either consider the virial temperature when it was relevant to the question of shock heating - at the time of accretion - or restrict the analysis window to a short enough time-scale over which the virial temperature does not appreciably change.

\subsection{Integrated Baryonic Budget}

\begin{figure*}
\centerline{\includegraphics[angle=0,width=5.8in]{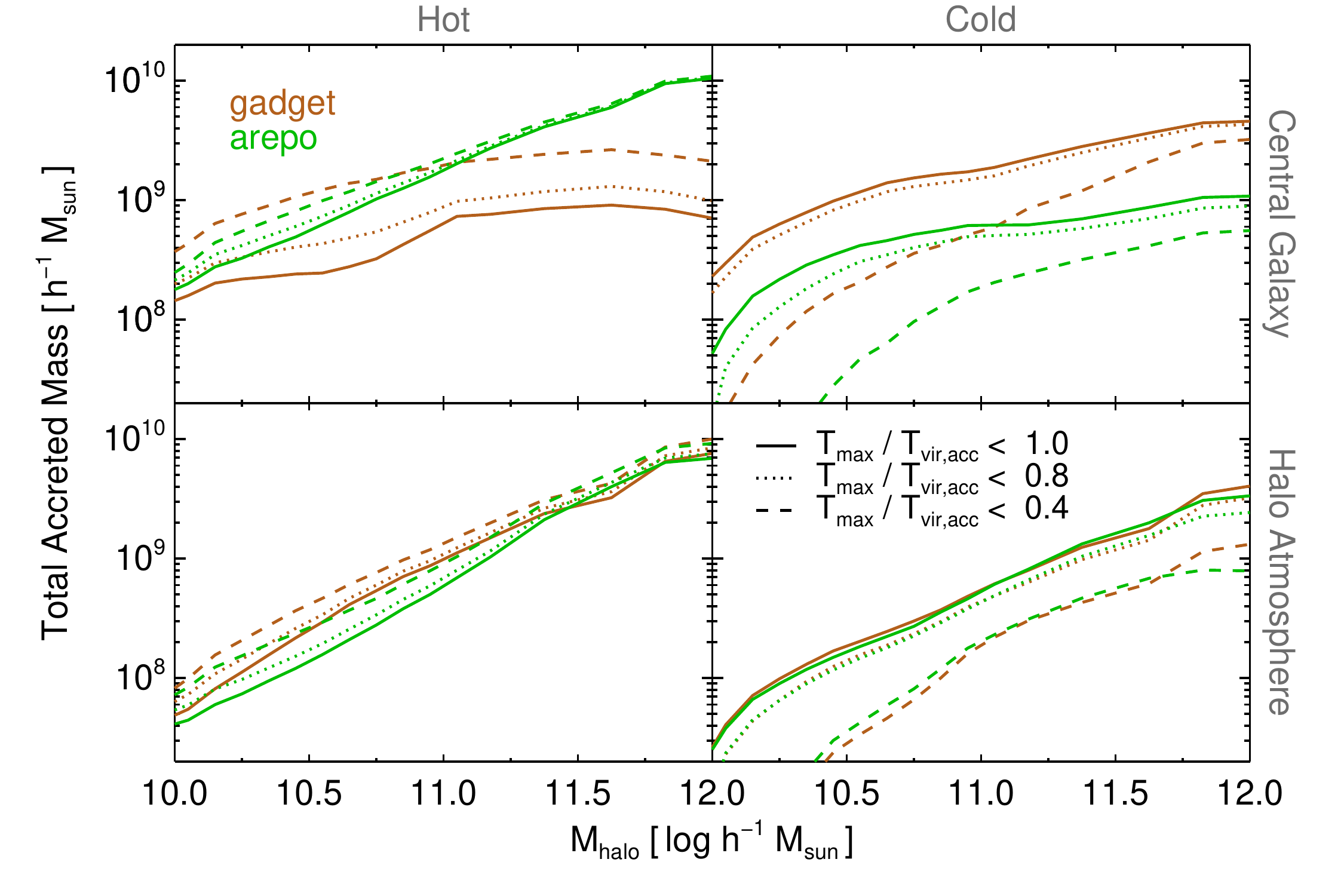}}
\caption{ The total amount of gas smoothly accreted onto either galaxies or halo atmospheres by $z=2$ decomposed into hot and cold contributions. Following our earlier discussion we make this separation based on some ratio of $T_{\rm max}$ to $T_{\rm vir,acc}$, as shown. Unlike in previous figures, we include gas accreted over an ``integrated'' time window encompassing the evolution of these systems back to $z=6$ or as far as they have well-defined parents.
 \label{fig_totalmass}} 
\end{figure*}

The strongly discrepant history of gas which eventually builds up galaxies by $z=2$ has important ramifications for the morphology and structure of these forming galaxies. In Figure \ref{fig_totalmass} we show the total gas mass smoothly accreted into both galaxies (top) and halo atmospheres (bottom), decomposed into hot and cold contributions based on a comparison of $T_{\rm max}$ to $T_{\rm vir,acc}$. Here, in contrast to earlier figures, we extend the ``time window'' over which accretion is counted to include a longer evolutionary history of these systems. We extend back in time as far as each halo is successfully tracked in our parent tree scheme, up to $z=6$. Consequently Figure \ref{fig_totalmass} shows the integrated contribution of smoothly accreted baryons to the $z=2$ mass budgets of these systems. When compared to the accretion rates calculated only over the past $1$ Gyr we find that the previous conclusions regarding the comparisons between the two codes and the balance between the hot and cold accretion modes remain qualitatively the same. Since the two simulations have the same initial conditions, we naturally expect their properties to tend towards better agreement at sufficiently high redshift. In particular, comparing the total accreted mass onto galaxies at halo masses of $\simeq$10$^{12}$\msun\, we find that {\small AREPO} has a factor of $\simeq$10 higher mass accreted with $T_{\rm max} > T_{\rm vir}$, and a factor of $\simeq$3 lower mass accreted with $T_{\rm max} < T_{\rm vir}$ when compared to {\small GADGET}. This represents a significant discrepancy in the gas accretion history integrated down to $z=2$, which is directly related to the morphological and kinematic differences of the respective galaxy populations.

A long standing difficulty of SPH has been its ability to produce central objects with angular momentum sufficient to form rotationally supported disks with realistic scale lengths \citep{vog12}. The relative differences between the angular momentum content of cold gas accretion in {\small GADGET} when compared to hot gas accretion in {\small AREPO} explains why galaxy properties differ between these two approaches. {\small AREPO} galaxies are more disk-like and more rotationally supported, with systematically larger disk scale lengths \citep{torrey12} in part because they are significantly built from the diffuse, hot halo gas, which has a large reservoir of angular momentum \citep[see also][]{sijacki12}. In contrast, {\small GADGET} galaxies are more compact and less rotationally supported in part because they are being built mainly from the cold gas, which is clumpy due to numerical artifacts and which tends to artificially lose its angular momentum. That realistic disk morphologies are related to the angular momentum reservoir from which they form has also been found regardless of numerical technique, in the context of galactic fountain material \citep{brook12} and cooling from hot coronae \citep{sales12}. Here the differences in the hot halo itself may also play a role. As we discuss below, hot halo gas in {\small AREPO} has a smaller spatial extent and reaches somewhat higher central densities than in {\small GADGET} \citep{keres12}, which leads to shorter cooling times and thus larger hot accretion rates onto centrally forming galaxies. 

\subsection{Geometry of Accretion}

\begin{figure*}
\centerline{\includegraphics[angle=0,width=5.5in]{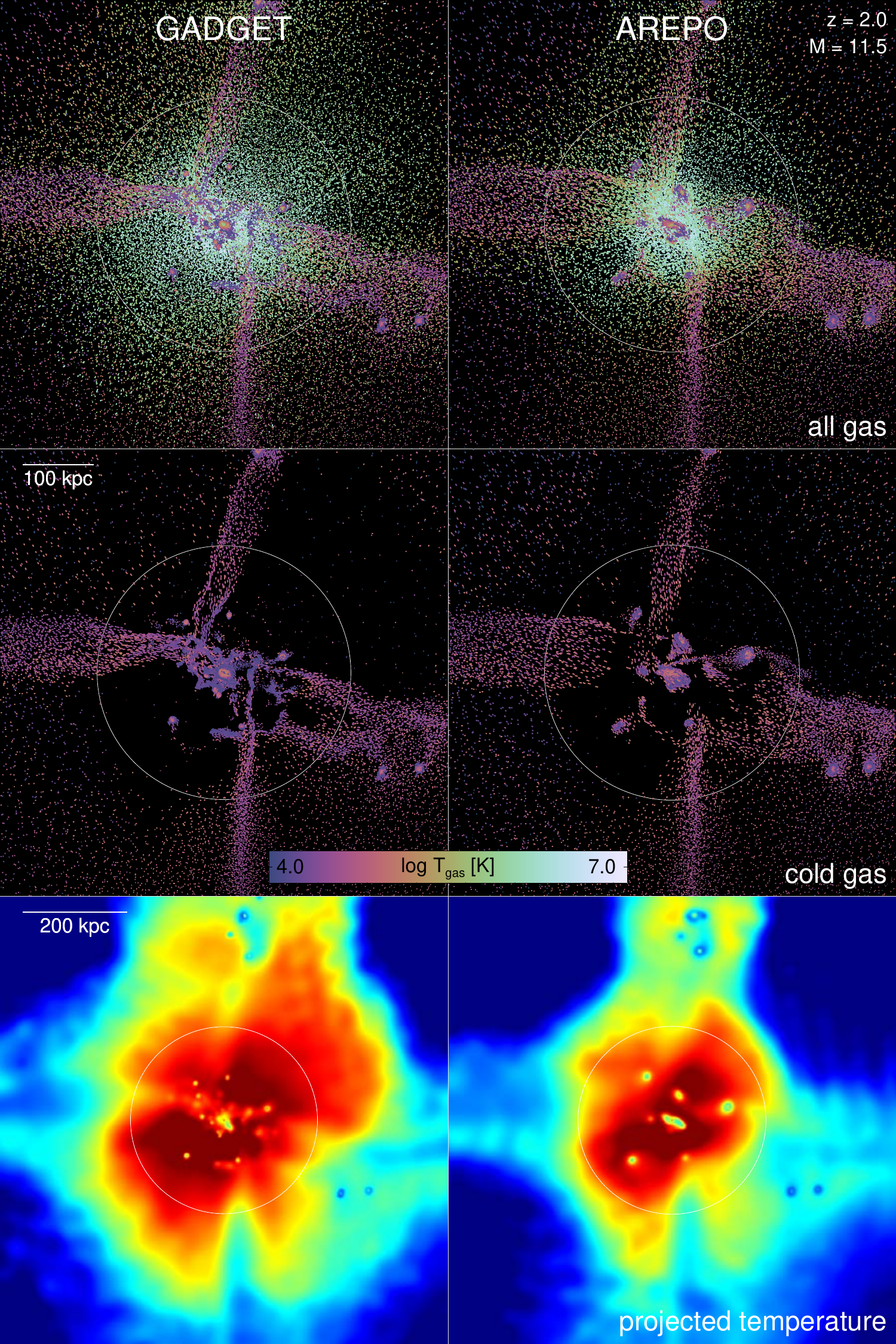}}
\caption{ Density and temperature distribution of individual gas elements for a $\simeq 10^{11.5}$\msun \, halo at $z=2$ in {\small GADGET} (left panels) and {\small AREPO} (right panels). The top panels show gas of all temperatures, while the middle panels show only gas with instantaneous temperature $T < 10^5$ K. Ticks are color coded by instantaneous temperature and with directions representative of the local velocity field. The bottom panels show the mass-weighted temperature projection for lines of sight through a larger cube of side length $5 r_{\rm vir}$. The same dark matter halo in {\small AREPO} has a less extended hot component than its {\small GADGET} counterpart. In both cases locally overdense and overcool filamentary gas structures penetrate the hot halo at $r_{\rm vir}$ and deliver gas to smaller radii. In {\small GADGET} these filaments become extremely collimated and cold at $\sim$0.5$r_{\rm vir}$ whereas in {\small AREPO} they tend to become more diffuse, dissipate, or experience significant heating. This last case can be seen at the top of the left filament which shows a thin, $\sim 10^6$ K streamer. 
 \label{fig_scattermaps}} 
\end{figure*}

\begin{figure*}
\centerline{\includegraphics[angle=0,width=5.5in]{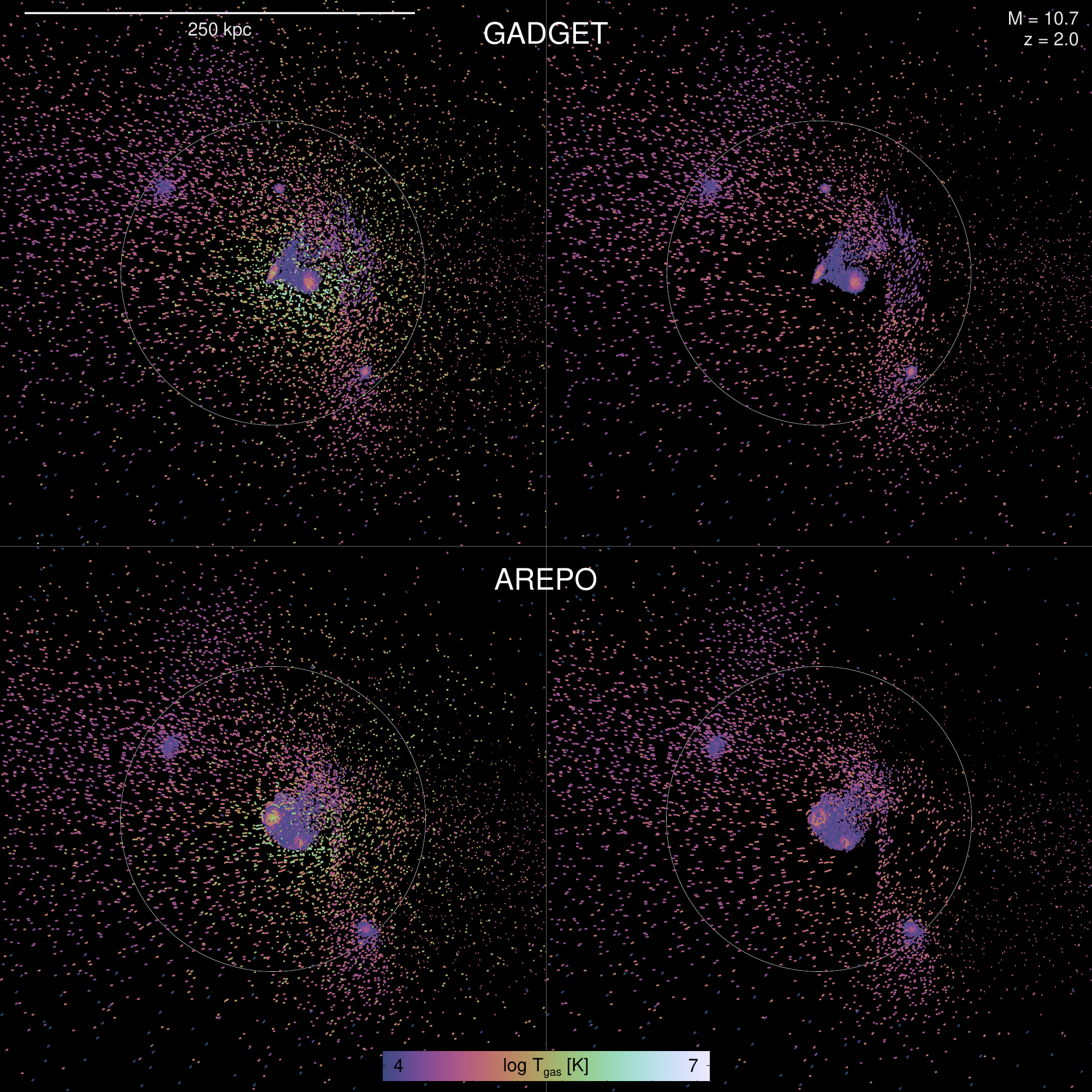}}
\caption{ Density of individual gas elements for a $\simeq 10^{10.7}$\msun \, halo at $z=2$ ($T_{\rm vir} \simeq 10^{5.4}$ K) in {\small GADGET} (top panels) and {\small AREPO} (bottom panels) color coded by instantaneous temperature and with directions representative of the local velocity field. The left panels show gas of all temperatures, while the right panels show only gas with instantaneous temperature $T < 10^5$ K. This halo, as with a majority of lower mass systems, does not show the same prominence of filamentary gas inflows.
 \label{fig_scattermaps2}} 
\end{figure*}

\begin{figure*}
\centerline{\includegraphics[angle=0,width=7.0in]{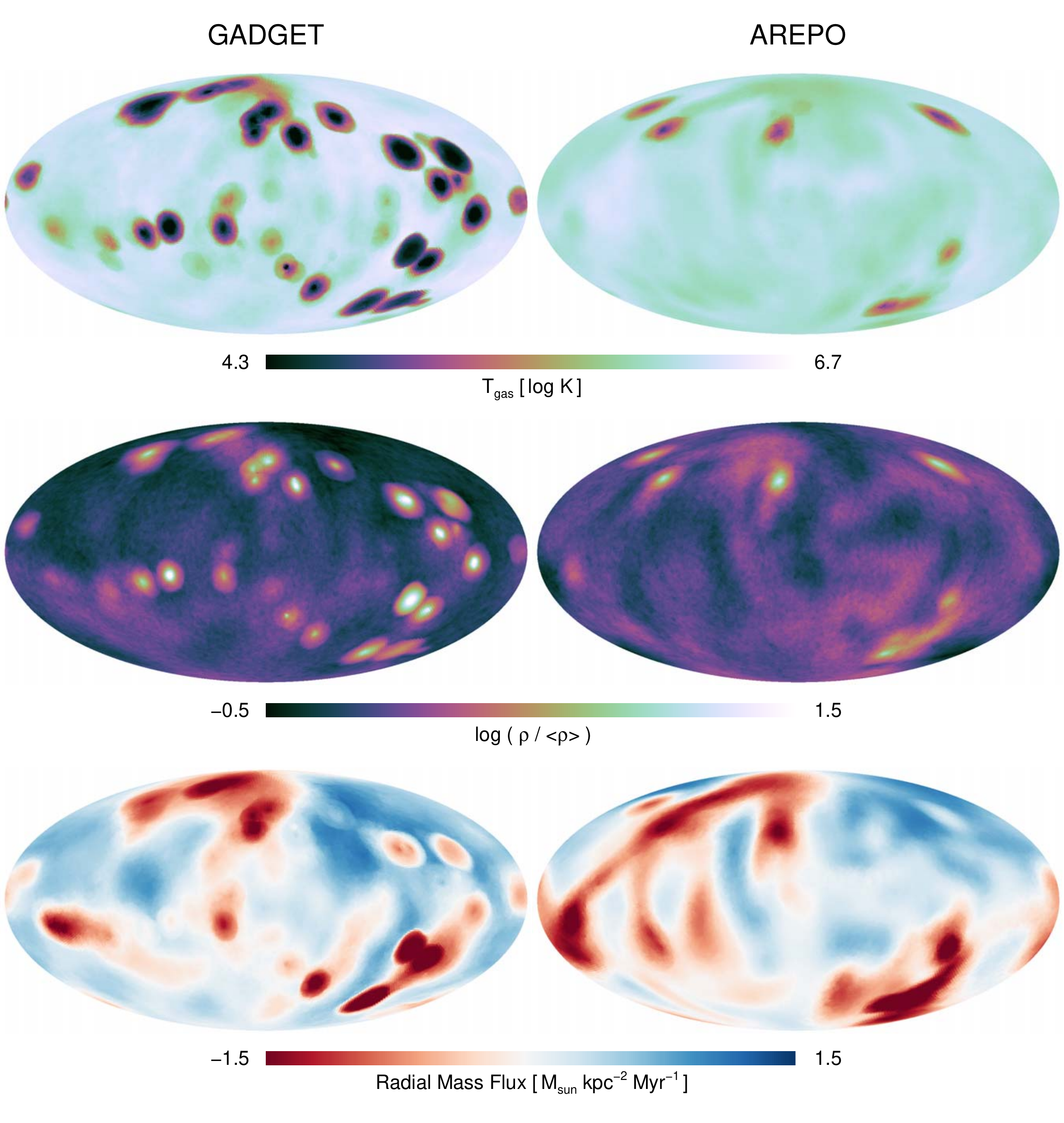}}
\caption{ An all-sky Mollweide projection of temperature, overdensity, and radial mass flux on the surface of the half virial sphere centered on one halo with $\log(M) \simeq 11.8$. All gravitationally bound substructures have been removed. The clustered regions of negative radial mass flux represent cross sectional views of gas filaments. They are associated with mass overdensities and cold temperatures with respect to the shell averages. The larger arcing features in the map are infalling sheets of gas with large angular extent on the sky.
 \label{fig_shells}}
\end{figure*}

\begin{figure*}
\centerline{\includegraphics[angle=0,width=7.0in]{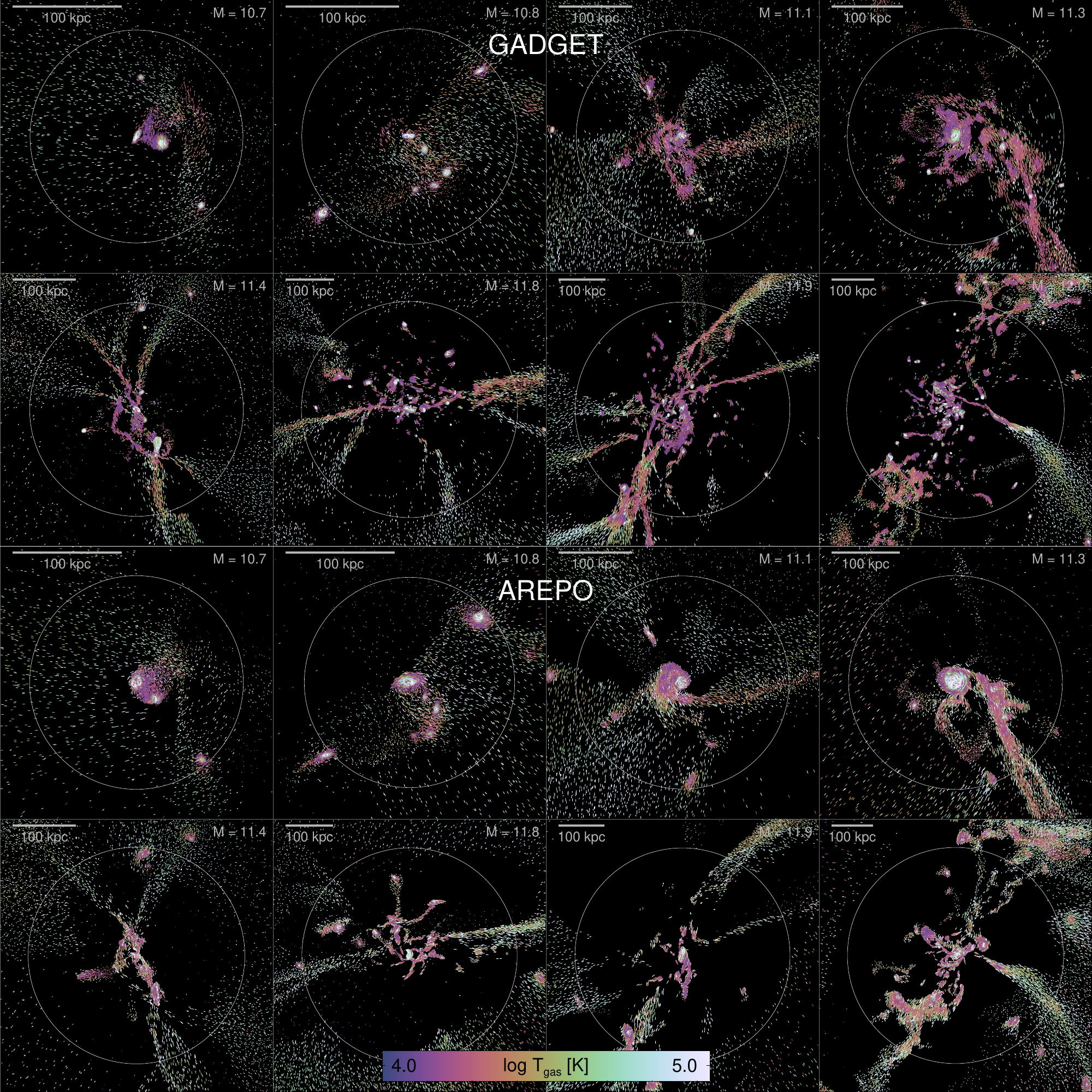}}
\caption{ Comparison of strictly cold gas with instantaneous temperature below $10^5$ K for eight matched systems at redshift two. The top eight panels show systems from the {\small GADGET} simulation while the bottom eight panels show the identical systems in {\small AREPO}. We commonly find examples of predominantly spherical accretion in low halo masses and differences in filamentary gas structures for larger halo masses, as discussed in the text.
 \label{fig_mosaic}} 
\end{figure*}

We find that the filamentary nature (geometry) of accreting gas at and around the virial radius of massive haloes is in good qualitative agreement between {\small GADGET} and {\small AREPO}. At high redshift ($z=2$) filaments are a common feature in a majority of massive ($M \geq 10^{11.5}$\msun) haloes. Prior to this section, we have included only smoothly accreted gas in our analysis and discussion. When visualizing the geometry of accretion, we show all gas elements, and do not restrict to only the smooth accretion mode as in previous plots. Figure \ref{fig_scattermaps}\ shows a gas projection of a particular halo matched between the two simulations, where in the top two rows temperature is mapped to particle color and the local velocity field by tick direction. The top panels shows all temperature gas, while the middle panels show only gas with instantaneous temperature below $10^5$ K. In {\small GADGET} (left column) we observe that filamentary gas structures often coherently flow to $r \leq 0.25 r_{\rm vir}$ and become progressively smaller in cross section. They can connect directly to central gas structures at $r \leq 0.1 r_{\rm vir}$, although the contribution of these thin streams to the total accretion rate of gas onto the galaxy is unclear. Their temperature profiles remain well below that of the virialised halo gas and show little evidence for shock heating. In {\small AREPO} (right column) these same gas filaments are more diffuse and experience significant heating at comparable radii. For instance, the two filaments from the top and left which seem to ``disappear'' in the right panel in fact heat to the temperature of the halo gas while remaining coherent. We discuss this important contribution to the accretion rates of hot gas below.

Another significant difference between the two results is the distribution of the hot halo gas itself. The bottom row of Figure \ref{fig_scattermaps} shows the mass-weighted temperature projection for the same halo in both {\small AREPO} and {\small GADGET}. In order to make a fair visual comparison we integrate each sightline by spatially distributing the gas mass of both SPH particles and Voronoi cells using the same SPH kernel approach. We observe that hot halo gas in {\small GADGET} extends to larger radii than in {\small AREPO}. To quantify this effect, we measure the stacked radial temperature profile of all haloes in the box with $10^{11.5}$\msun$ < M_{\rm halo} < 10^{12.0}$\msun. At the virial radius, the gas temperature in {\small AREPO} haloes is a factor of $\simeq 2/3$ lower than in {\small GADGET} -- the mean gas temperature at $r_{\rm vir}$ in {\small GADGET} is obtained at $\simeq 0.7 r_{\rm vir}$ in {\small AREPO}. As we discuss in the context of accretion rates, the smaller hot gas reservoir in {\small AREPO} haloes results from more efficient cooling out of the halo, while in the case of {\small GADGET} this cooling is suppressed due to spurious numerical heating.

An independent point which has led to confusion is the nature of accretion in lower mass haloes. Figure \ref{fig_scattermaps2}\ shows an example of such a system with M$_{\rm halo} \simeq 10^{10.7}$\msun\, at redshift two. In this case there are no prominent filaments, nor do we expect any \citep{katz03,keres09}. The virial temperature of this system is $\simeq$10$^{5.4}$ K and would lie at the low mass end of the transition mass region from cold to hot mode dominated. Given the usual definition of $\log T_c=5.5$, gas which had accreted and shock heated to the virial temperature would be considered cold mode accretion. In contrast, under a definition based on some fraction of $T_{\rm vir}$ a large fraction of that same gas would be considered hot mode. Indeed, for this mass range in Figure \ref{fig_coldfrac_var}\ we expect that approximately half of the halo gas has $T_{\rm max} > T_{\rm vir}$ and half has $T_{\rm max} < T_{\rm vir}$. It is then critical to differentiate between gas which shocks to $T_{\rm vir} \leq T_c$ and gas which accretes ``cold'' into low mass haloes with short cooling times. This is particularly the case since the signature of both mechanisms is similar in maximum past temperature, making any separation based on temperature alone difficult if not impossible.

In order to better compare the properties of filaments we construct maps of various fluid quantities which reveal the cross-sectional profiles of these structures. Figure \ref{fig_shells}\ shows gas temperature, density and radial mass flux calculated on the surface of a sphere centered on one particular halo, the $\log(M)=11.8$ example from Figure \ref{fig_mosaic}, and with a radius equal to half the virial radius. We use a simple mass-weighted tophat kernel to interpolate these quantities onto equal area pixels using the {\small HEALPIX} scheme \citep{gorski05}, and display the result using a Mollweide all-sky projection.

Pronounced differences appear in images of projected gas density. We find that the radial mass flux inward at these radii is dominated in the SPH halos by a large number of thin filaments and gas blobs with progressively smaller cross section. The moving mesh case shows no evidence of gas structures of comparably small angular extent. The density profiles of the filaments are also not as centrally concentrated and have larger angular extent. Although filaments reach roughly the same minimum temperature, the transition to the hot $10^{5.5}-10^6$ K gas in the halo is more gradual in {\small AREPO}. That is, filaments have less well-defined boundaries, indicative of increased mixing, in the moving mesh calculation, though they appear to be roughly at the same level of pressure equilibrium. Although the angular covering factor of infalling material is slightly larger in {\small AREPO}, it is interesting that for this particular halo, where we have accretion onto the central galaxy dominated by $T_{\rm max} > T_{\rm vir,acc}$ gas, the mass flow is still highly aspherical. That these filaments of hot gas, originating from large scale features in the IGM, contribute to the hot gas accretion rate distinctly from classical cooling flows out of hot halo gas is the third key point of this work. The presence of this coherent hot accretion implies that a simple cooling rate calculation based on halo properties cannot accurately describe high redshift accretion in massive haloes. This may have important consequence for semi-analytical models (SAMs) which assume radiative cooling from a quasi-static atmosphere.

Comparing large $\simeq$\,$1$\,Mpc scale gas and dark matter density fields, the cold $\simeq$\,$10^4$\,K filaments which are relatively narrow within the virial radius can clearly be identified as extensions of filaments of larger diameter outside the halo. Given that {\small GADGET} and {\small AREPO} share the same treatment of dark matter and the same gravity solver, and show good agreement in IGM properties outside haloes \citep{bird12}, it is unsurprising that at the virial radius the differences we observe above minimise and the two codes show similar gas properties.

We show eight further examples of strictly cold gas below $10^5$\,K in matched systems spanning the mass range $10.7 \leq \log(M_{\rm halo} [M_{\odot}]) \leq 12.0$ in Figure \ref{fig_mosaic}\ illustrating that the similarities and differences we describe are commonly found in the simulated cosmological volume\footnote{A catalog of similar visual comparisons for all objects with $M > 10^{10.75}$\msun\, at $z=\{0,1,2,3\}$ is available online at \texttt{www.cfa.harvard.edu/itc/research/movingmeshcosmology/}.}. Haloes in both simulations support approximately the same frequency of filamentary structures over any given range of halo mass. The exception appears to be filaments that are tidal in nature, which are much more pronounced -- longer and more concentrated -- in {\small AREPO}. The cold gas distribution in the most massive {\small GADGET} haloes is dominated by the large blob populations, which are absent in the corresponding {\small AREPO} systems.

\subsection{The Clumpy Contribution}

\begin{figure*}
\centerline{\includegraphics[angle=0,width=5.8in]{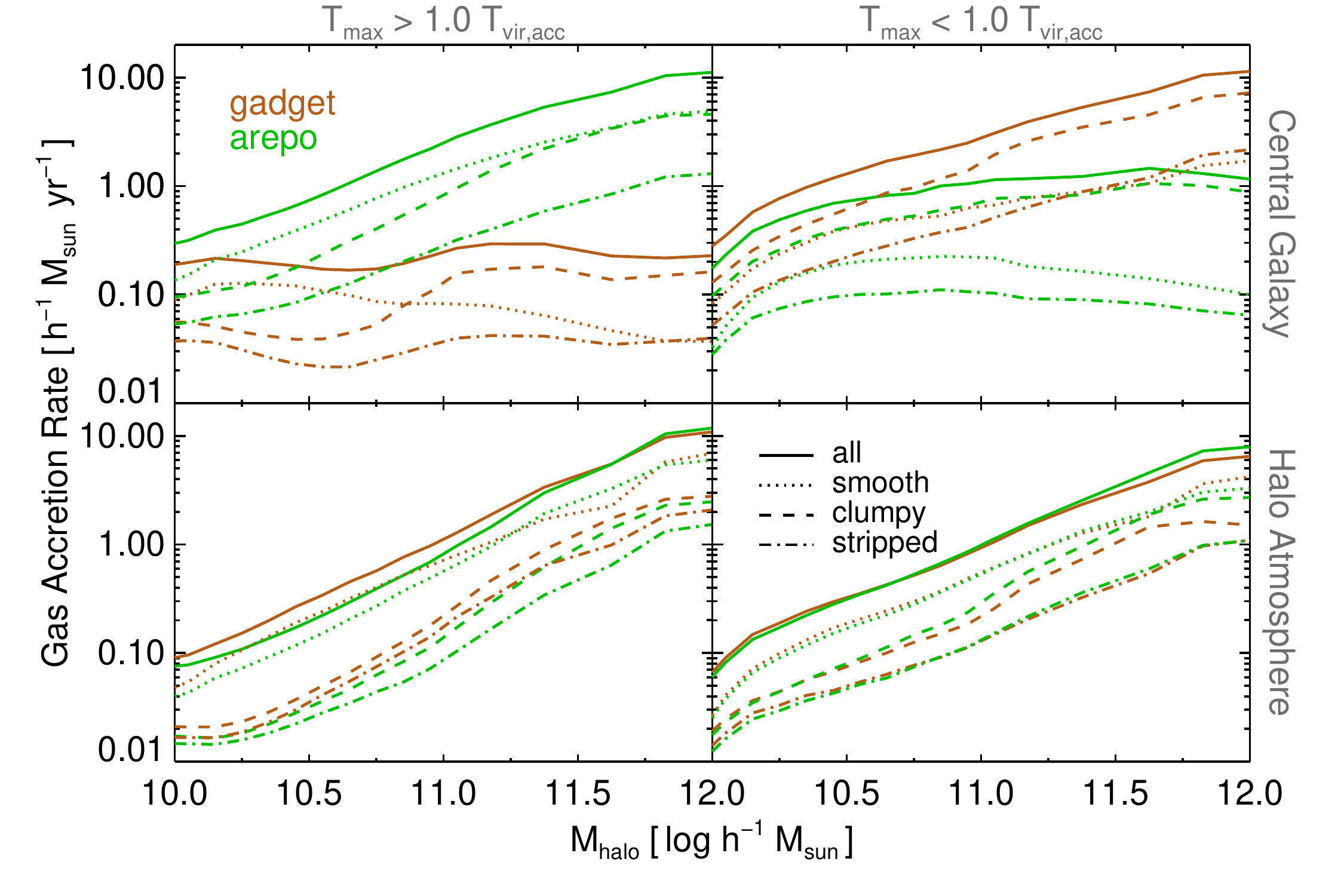}}
\caption{ The total accretion rate separated into contributions from our three different modes: smooth (dotted), clumpy (dashed), and stripped (dot-dashed).
 \label{fig_accRateMode}} 
\end{figure*}

We seek to understand the accretion of material onto haloes and galaxies originating directly from the intergalactic medium. If we include all material, the resulting cold fractions and cold mode accretion rates are only upper limits, due to some nonzero contribution from minor mergers. Past numerical simulations run with SPH and focusing on cold mode accretion have specifically addressed this issue of merging substructures, both resolved and unresolved \citep{keres05,keres09,brooks09,vdv11a}. The exact approach in separating out the merger contribution differs. In \cite{brooks09} and in this study we remove material that belonged to any galaxy halo other than the main galaxy under consideration, while in \cite{keres09} for instance only the central galaxy and not its associated circumgalactic (or halo) material is removed. Other studies have neglected this distinction and included the accretion of dwarf satellites as part of a cold mode \citep{shen12}. Although simulations using the AMR technique note the presence of substructure embedded within filaments, they also neglect to make this separation \citep{dekel09,agertz09} though due to technical constraints. One option as in \cite{ocvirk08} is to use an instantaneous criterion such as gas density to identify a clumpy component. However, without some additional tool for tracing gas properties back in time (tracer particles) the contribution of substructures and associated material cannot be unambiguously accounted for, and a measurement of the cold gas accretion from the IGM represents an upper limit.

Applying the three definitions for ``smooth'', ``clumpy'', and ``stripped'' from Section \ref{ssMeasAcc} we decompose the accretion rates separately for hot and cold accretion in Figure \ref{fig_accRateMode}, where we here use $1.0 \times T_{\rm vir,acc}$ as the separating criterion. As previously discussed, our definition of clumpy includes all material bound to satellites -- both galaxies and their associated circumgalactic medium. We focus first on cold material accreted onto central galaxies, for which the clumpy mode is the $\simeq 60\%$ dominant contributor at all halo masses in both {\small GADGET} and {\small AREPO}. Smooth accretion accounts for only $\simeq 10\%$ of the cold mass accumulation at the massive end, where most of the cold accretion onto central galaxies arises from merging substructure. For the central galaxies of the lowest mass systems we consider with $\simeq 10^{10}$\msun\, the smooth contribution is slightly larger ($\simeq 20\%$). This fraction is relatively converged with resolution; however,  any estimate of the merger contribution for the lowest mass systems may be underestimated due to numerical resolution. For all halo masses the stripped component is comparable to the smooth component. The accretion of hot material onto central galaxies shows similar behavior, except that the contribution from clumpy mode gas scales up with increasing halo mass, while the contribution of smooth gas scales down. The only notable difference between the two codes is for material accreted hot onto galaxies hosted in massive haloes with $M_{\rm halo} \geq 10^{11}$\msun\, for which {\small AREPO} shows a lower clumpy fraction ($40\%$ vs $70\%$) and a higher smooth fraction ($40\%$ vs $20\%$), consistent with the picture that smooth gas is more efficiently heated in {\small AREPO}.

The importance in separating different gas origins in the simulations can be seen by reference to observations of our local environment, in which the Magellanic Stream is a large reservoir of cold, neutral gas thought to arise from the tidal interaction of dwarf satellites \citep{besla10,besla12}. Its significant spatial extent through the halo of the Milky Way does not arise from a gas filament of direct cosmological origin, although it might easily be characterised as such. Similar features in external galaxies will be even more difficult to characterise. Although separating the contribution of minor mergers and accreted tidal debris from gas infalling directly from the IGM presents a technical challenge for simulations, it is necessary to truly identify the origin of the gas that ends up forming galaxies, without mixing up contributions from disparate physical origins.

\subsection{Resolution Convergence}

\begin{figure*}
\centerline{\includegraphics[angle=0,width=5.8in]{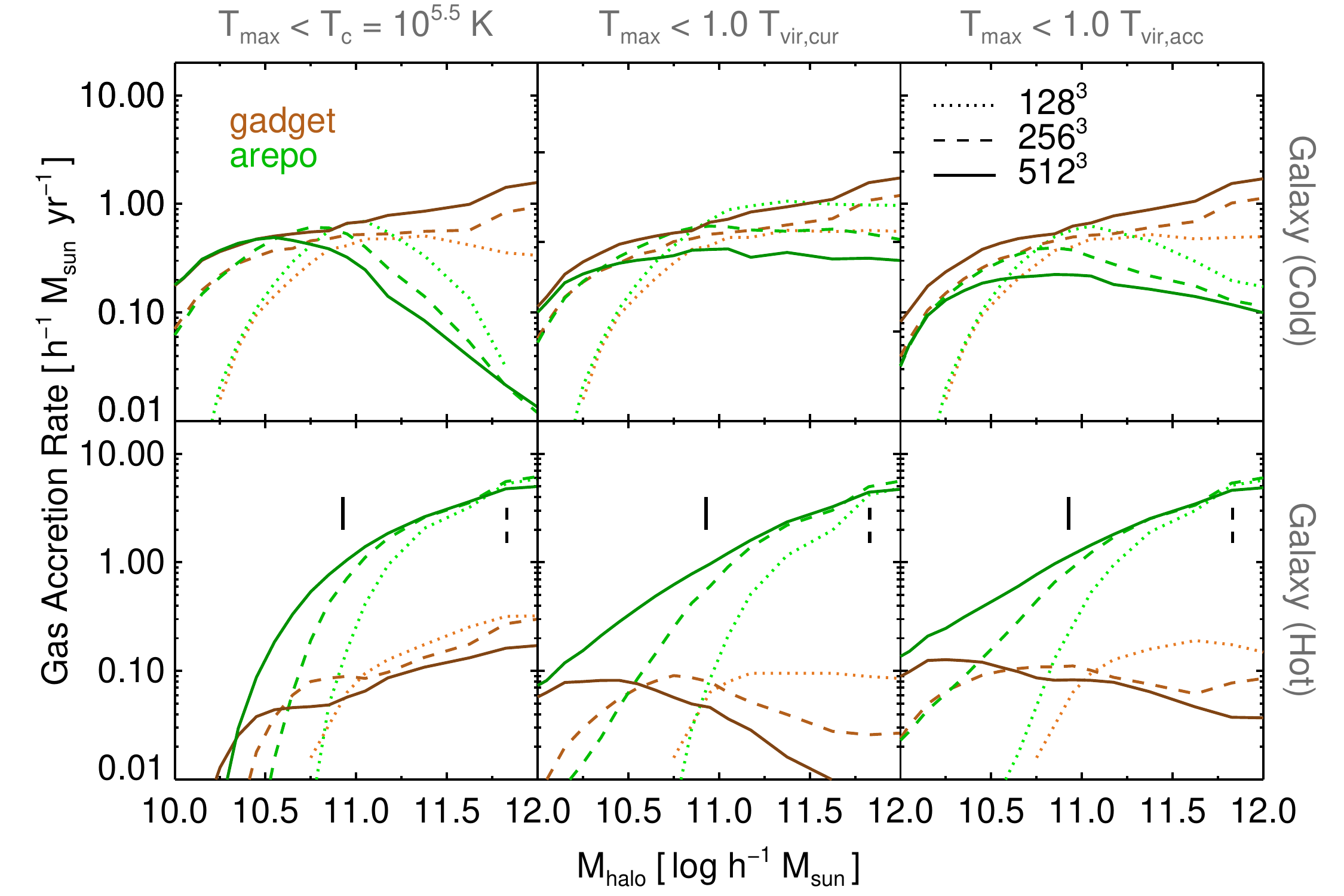}}
\caption{ Convergence of the accretion rate onto galaxies with resolution. The vertical black lines show the halo masses at which $\Omega_b M_{\rm halo} / m_{\rm target} \simeq 5 \times 10^4$ for the $512^3$ and $256^3$ runs, a rough requirement on the number of gas mass elements for which gas accretion physics begins to be adequately resolved and converged in {\small AREPO}.
 \label{fig_accRateRes}} 
\end{figure*}

In this last section we address some of the numerical issues related to our moving mesh calculation. In Figure \ref{fig_accRateRes}\ we show the convergence of the gas accretion rate onto central galaxies for our three resolution levels. The total (the sum of the top and bottom panels for any given column) is well converged across the given halo mass range, and all resolution trends are qualitatively similar regardless of the separating line drawn between hot and cold. In general, \cite{vog12} showed the increase in high redshift global star formation rates due to better resolving dense gas in low mass haloes. We expect that smooth accretion rates will then decrease as resolution increases, with a corresponding increase in either clumpy or stripped accretion, as more gas mass collapses into substructures at early times. Galaxies in {\small AREPO} exhibit this behavior for material accreted cold, i.e. directly from the IGM. In contrast, the accretion rate of hot material from the halo increases strongly with resolution for halo masses which are poorly resolved in our simulations. This enhanced cooling from the halo occurs due to a combination of the changing temperature distribution of hot halo gas, together with inadequately resolved mixing in two particular regimes, as we discuss below.

First, \cite{keres12} shows that the radial temperature profiles of {\small AREPO} haloes become steeper and reach higher maximum temperatures with increasing resolution. We verify this and furthermore find that it is particularly problematic for poorly resolved systems where the number of gas cells in the entire halo is of order a few thousand or fewer. For more well-resolved haloes ($M_{\rm halo} \geq 10^{11}$\msun) the temperature profiles are converged between our medium and high resolution runs. This occurs in part because the halo material more fully populates the high temperature tail of the thermal distribution of the virialised gas.

The strong mixing between the rotating galactic disk and the halo gas identified by \cite{sijacki12} is also artificially enhanced at low resolution where this interface is poorly resolved. This shortens the cooling times of hot halo gas near the disk and drags down the temperature profile, which ultimately results in accreting material heating to a lower peak temperature. The potential for overmixing at low resolution is a deficiency inherent to grid codes, including {\small AREPO}. In this case the moving mesh nature of the scheme acts to minimise the effect, though it cannot be suppressed entirely as in SPH codes. The same issue arising at the disk-halo interface also occurs in a second regime, with infalling substructures which penetrate into the halo and lower the cooling times of the hot gas due to overmixing. This is in contrast to SPH, where a relatively poor treatment of fluid instabilities prevents efficient stripping of gas from infalling satellites \citep{sijacki12}.

Galactic disks in the low mass haloes are also poorly resolved in our lowest resolution simulations \citep{keres12}, which may extend them beyond our $0.15 r_{\rm vir}$ cut, particularly in {\small AREPO} where they are larger and more extended to start with. The result is an underestimation of accretion rates, while at higher resolution these disks are more compact and the accretion is largely recovered. This effect may preferentially influence gas with larger $T_{\rm max}$ which is found at systematically larger radii and lower density in galactic disks.


\section{Discussion and Conclusions} \label{sDiscussion}

We are motivated to define a cold accretion mode based on the markedly distinct channel of cold filamentary streams penetrating the hot atmospheres of massive haloes. Given the existence of two distinct modes, how can we distinguish between the two? We observe that this form of gas accretion enables gas to infall onto galaxies without shock heating to an appreciable fraction of the virial temperature. This motivates a selection criterion based on virial shocking, or lack thereof. This is not a new idea. Indeed, \cite{keres05} explored this option but concluded that the distinction between hot and cold was less clear than with a constant temperature cut. \cite{vdv11a} likewise compared hot fractions obtained using either a threshold $\log T_c=5.5$ or various fractions of $T_{\rm max}/T_{\rm vir}$. That work concluded, as we do, that the hot fraction depends strongly on the choice of criterion. Furthermore, \cite{vdv11a} acknowledges that gas accreted in their ``cold mode'' may have in fact gone through a virial shock, indicating that it is critical to investigate the gas selection made with a global constant temperature threshold over the range of halo masses being considered.

Furthermore, even comparing the temperature history of gas elements to the virial temperatures of their parent haloes is an inexact probe of whether or not gas shock heats. The picture that gas shock heats up to $T_{\rm vir}$ at $r_{\rm vir}$ is an over simplification of a virialisation process that may involve multiple shocks at increasingly higher temperatures as infalling material encounters the temperature gradient of halo gas. Indeed, shock heating may not even be the most fundamental question as to how galaxies obtain their gas. This might be better answered by considering the cooling time of gas \citep{wr78,wf91}, and so the time-scale for collapse onto a forming galaxy. The radial stalling associated with incorporation into the hot halo gas may be more useful than the gas temperature itself. In the same manner, the angular momentum history of infalling gas may provide a more physically motivated segregation of different accretion modes. We have found that the interpretation of the angular momentum, for instance, is not nearly as clear as with maximum past temperature and we suggest this is the primary reason why $T_{\rm max}$ remains the predominant technique used to study cold mode accretion.

Based on additional temperature cuts, recent simulations have also identified a third ``warm'' mode of gas accretion \citep{joung12,murante12} in the $10^5$\,K $<$ T $<\,10^6$ K regime and selected by constant temperature bounds over either instantaneous temperature (in \cite{joung12} using AMR) or maximum past temperature (in \cite{murante12} using SPH). In the case of \cite{murante12} this prominent mode is attributed to effective SN thermal feedback. Unfortunately, since these studies are focused on low redshift and include galactic outflows we cannot make a direct comparison to our results. We have not therefore investigated this particular temperature selection criterion, although a quantitative comparison with AMR results and an exploration of the impact of feedback are crucial and natural directions for future work. The commonly perceived agreement between SPH and AMR simulations, of cold mode accretion as a dominant contributor to galaxy formation, is based on a mixed analysis of instantaneous gas properties and qualitative comparisons. An as of yet unperformed comparison of the thermal history of accreted gas (using tracer particles) is required to demonstrate quantitative agreement between these two methods.

\subsection{Numerical Uncertainties and Included Physics}

Particularly when studying gas accretion, a perceived strength of SPH, embodied in codes such as {\small GADGET}, is what should accurately be termed its pseudo-Lagrangian nature \citep{vog12}. The consequence of this behavior is the apparent ease by which the thermodynamic properties of individual mass elements can be traced through time. However, consider the example of mass represented by a single SPH particle in a shearing flow, as illustrated in Fig. 20 of \cite{vog12}. This mass is initially drawn from a volume centered on the SPH particle with radius of order the local smoothing length. Shearing of this volume deforms its shape and the associated mass should follow. However, the nature of SPH is such that this mass is forced to remain bound to its SPH particle. The consequence is that the evolution of the fluid is not fully consistent with the equations of motion and cannot then be truly Lagrangian. In this regard {\small AREPO} is also not strictly Lagrangian since the gas cells are not allowed to become arbitrarily distorted. However, the fundamental difference is that mass exchange between cells occurs in a manner consistent with the equation of mass conservation and we can therefore term the scheme quasi-Lagrangian, because unlike in SPH the solution is faithful to the underlying equations of motion.

It is also important to note the differences arising due solely to different formulations of SPH. \cite{keres05} used {\small TreeSPH} \citep{hk89} and found a strongly bimodal histogram of the maximum past temperature of accreted gas, and comparable contributions of the hot and cold modes. \cite{keres09} used {\small GADGET-2} and concluded that at $z \geq 2$ the accretion is dominated by the cold accretion mode at all halo masses, with little evidence for bimodality in the same $T_{\rm max}$ histogram. The original motivation for a constant threshold temperature used to separate hot and cold modes is in fact the location of the temperature minimum resulting from this bimodality. Yet, this same constant value has been used in virtually all simulation based investigations of cold mode accretion since. 

\cite{yoshida02} compared these two SPH formulations and concluded that the {\small TreeSPH} technique leads to a significant over-cooling stemming from its geometrical symmetrization of hydrodynamical forces. In the context of cooling from halo gas this problem manifests as increased hot mode accretion. Accordingly, \cite{keres09} attributes the much higher hot mode accretion rates in their earlier simulations to numerical deficiencies in the older SPH formulation. The larger relative contribution of high $T_{\rm max}$ gas and the dominant contribution of hot gas accretion in the most massive haloes found in \cite{keres05} is in fact in better agreement with this work than that found in \cite{keres09}. However, this agreement is a coincidental result of the aforementioned numerical inaccuracies in {\small TreeSPH}.

It is not altogether surprising, then, that our moving mesh results with {\small AREPO} are in tension with previous studies, particularly SPH simulations made using the standard formulation found in {\small GADGET}. Significant uncertainties remain in the numerical simulation of cosmological volumes where the inclusion of a diverse number of physical processes over a large dynamic range in spatial and temporal scales is both computationally intractable and yet phenomenologically required. What is surprising is the extent to which differences arise solely due to the numerical scheme employed to solve the hydrodynamics, and not to differences in subgrid prescriptions, star formation recipes, feedback implementations, and the like. The simulations presented in this work implement a basic set of physical processes in addition to gravity and hydrodynamics. We leverage this relative simplicity to make a clean comparison between the SPH and moving mesh techniques. However, it is probable that strong feedback from stars or active galactic nuclei could significantly alter the nature of gas inflow from the IGM, and this remains an open question for future work. The moving mesh implementations of ideal magnetohydrodynamics \citep{pakmor11} and physical viscosity \citep{munoz12} in {\small AREPO} will likewise enable us to investigate these potentially important effects for cosmological gas accretion.

\subsection{Conclusions}

We conclude with two main points. Firstly, that the constant temperature threshold commonly used to study cosmological gas accretion is only reasonable above some minimum halo mass, and application in an overly broad context biases conclusions regarding the relative importance of hot or cold gas accretion. Secondly, that numerical deficiencies inherent to smoothed particle hydrodynamics (SPH) simulations non-trivially modify the relative contributions of hot and cold mode accretion, under any definition. An identical analysis of gas accretion utilizing our Monte Carlo tracer scheme with the new moving mesh code {\small AREPO} demonstrates significant physical differences in the thermal history of accreted material. We summarise our primary results as:

\begin{itemize}
\item In agreement with previous work, {\small GADGET} simulations imply that at $z=2$ only a small fraction of gas in centrally forming galaxies of massive haloes above $10^{11}$\msun\, heats to an appreciable fraction of the virial temperature during accretion. In {\small AREPO} we find a decrease in the accretion rate of cold gas, by a factor of $\sim 2$ at $M_{\rm halo} \simeq 10^{11}$\msun, as well as a significantly enhanced accretion rate of hot gas, by an order of magnitude at the same halo mass. These discrepancies increase for more massive systems. We attribute the decrease in the cold accretion rate primarily to the large population of numerical ``blobs'' which efficiently deliver cold gas to central galaxies in our SPH simulations, but are completely absent in our moving mesh calculation. The increase in the hot accretion rate is dominated by significantly more efficient cooling from halo gas in {\small AREPO}, where spurious heating from the dissipation of turbulent energy on large scales prevents the same behavior in {\small GADGET}.

\item We argue that comparison of the maximum past temperature $T_{\rm max}$ of a gas element to a fixed temperature threshold $T_c$ makes physical sense only for haloes with $T_{\rm vir} \gg T_c$. For lower mass systems the past temperature history should instead be compared to some fraction of the virial temperature of the dark matter halo. However, at sufficiently low halo masses, when the virial temperature becomes comparable to the gas temperature in the IGM, neither a constant temperature threshold nor one scaled with the virial temperature are sufficient to probe gas shock heating and virialisation.

\item We observe that the ``transition mass'' from cold to hot dominated accretion which has been reported between $10^{11-11.5}$\msun\, is a consequence of the constant temperature criterion. When comparing the thermal history of gas instead to a fraction of $T_{\rm vir}$ we find no sharp transition, only a gradual decline from $f_{\rm cold} \simeq 20\%$ to $\simeq 0\%$ over the mass range from $M_{\rm halo} = 10^{10}$\msun\, to $10^{12}$\msun.
 
\item The filamentary geometry of accreting gas near the virial radius is a common feature of massive haloes above $M\sim 10^{11.5}$ at high redshift ($z=2$). Although characterised by the same large scale morphology, filamentary gas structures in {\small GADGET} tend to either remain cold and flow coherently to small radii within a halo, or artificially fragment and form a large number of ``blobs'' which are purely numerical in origin. In contrast, filaments in {\small AREPO} simulations are more diffuse and experience significant heating at comparable radii. The geometry and angular covering factor of material accreted with $T_{\rm max} \geq T_{\rm vir,acc}$ indicates that coherent, filamentary flows associated with large scale IGM filaments contribute significantly to hot accretion rates in these massive systems.

\end{itemize}


\textit{Acknowledgments.} DN would like to thank Claude-Andre {Faucher-Gigu{\`e}re}, Patrik Jonsson, Diego Mu$\tilde{\rm{n}}$oz, and Paul Torrey for many insightful discussions, comments, and suggestions. The computations presented in this paper were performed on the Odyssey cluster at Harvard University. DS acknowledges NASA Hubble Fellowship through grant HST-HF-51282.01-A.

\bibliographystyle{mn2e}
\bibliography{refs}

\end{document}